\documentclass[onecolumn]{aastex631}

\usepackage{xcolor}
\usepackage{upgreek}
\usepackage{comment}

\newcommand{\Poseidon}{\texttt{POSEIDON}}
\newcommand{\Brewster}{\texttt{Brewster}}
\newcommand{\Apollo}{\texttt{APOLLO}}
\newcommand{\change}{}

\graphicspath{{./}{figures/}}
\begin{document}

\title{Ross 458c: Gas Giant or Brown Dwarf?}

\author[0009-0001-6903-0131]{William W. Meynardie}
\affiliation{Department of Astronomy, University of Michigan, 1085 S. University Ave., Ann Arbor, MI 48109, USA}

\author[0000-0003-1227-3084]{Michael R. Meyer}
\affiliation{Department of Astronomy, University of Michigan, 1085 S. University Ave., Ann Arbor, MI 48109, USA}

\author[0000-0003-4816-3469]{Ryan J. MacDonald}
\altaffiliation{NHFP Sagan Fellow}
\affiliation{Department of Astronomy, University of Michigan, 1085 S. University Ave., Ann Arbor, MI 48109, USA}
\affiliation{School of Physics and Astronomy, University of St Andrews, North Haugh, St Andrews, KY16 9SS, UK}

\author[0000-0002-5335-0616]{Per Calissendorff}
\affiliation{Department of Astronomy, University of Michigan, 1085 S. University Ave., Ann Arbor, MI 48109, USA}

\author[0000-0003-0814-7923]{Elijah Mullens}
\affiliation{Department of Astronomy and Carl Sagan Institute, Cornell University, 122 Sciences Drive, Ithaca, NY 14853, USA}

\author[0009-0006-9292-3129]{Gabriel Munoz Zarazua}
\affiliation{Department of Physics and Astronomy, San Francisco State University, 1600 Holloway Ave., San Francisco, CA 94132, USA}

\author[0009-0003-9682-9533]{Anuranj Roy}
\affiliation{Department of Physics and Astronomy, San Francisco State University, 1600 Holloway Ave., San Francisco, CA 94132, USA}

\author[0009-0000-9583-0186]{Hansica Ganta}
\affiliation{Department of Physics and Astronomy, San Francisco State University, 1600 Holloway Ave., San Francisco, CA 94132, USA}

\author[0000-0003-4636-6676]{Eileen C. Gonzales}
\affil{Department of Physics and Astronomy, San Francisco State University, 1600 Holloway Ave., San Francisco, CA 94132, USA}

\author[0000-0002-7139-3695]{Arthur Adams}
\affiliation{Department of Astronomy, The University of Virginia, 530 McCormick Rd, Charlottesville, VA 22904, USA}

\author[0000-0002-8507-1304]{Nikole Lewis}
\affiliation{Department of Astronomy and Carl Sagan Institute, Cornell University, 122 Sciences Drive, Ithaca, NY 14853, USA}

\author[0000-0002-7006-0039]{Yucian Hong}
\affiliation{Data Scientist, New York City, NY, USA}%Machine Learning Engineer, Starr Insurance, New York, USA}%Department of Astronomy and Carl Sagan Institute, Cornell University, 122 Sciences Drive, Ithaca, NY 14853, USA}

\author[0000-0003-2279-4131]{Jonathan Lunine}
\affiliation{Jet Propulsion Laboratory, California Institute of Technology, Pasadena, CA 91109, USA}



\begin{abstract}

Ross 458c is a widely separated planetary mass companion at a distance of 1100 AU from its host binary, Ross 458AB. It is a member of a class of very low-mass companions at distances of hundreds to thousands of AU from their host stars. We aim to constrain Ross 458c’s formation history by fitting its near-IR spectrum to models to constrain its composition. If its composition is similar to its host star, we infer that it likely formed through turbulent fragmentation of the same molecular cloud that formed the host. If its composition is enhanced in heavy elements relative to the host, this lends evidence to formation in the disk and subsequent migration to its current separation. Here, we present high-resolution (R$\sim$2700) emission spectra of Ross 458c with JWST NIRSpec Fixed Slit in the F070LP, F100LP, and F170LP filters from 0.8 to 3.1 $\upmu$m. We fit these spectra using both grids of forward models (Sonora Bobcat, Sonora Elf Owl, and ExoREM) and atmospheric retrievals (\Poseidon). We also constrain the composition of Ross 458AB by fitting an archival SpeX spectrum with PHOENIX forward models. The forward model grids prefer an enhanced atmospheric metallicity for Ross 458c relative to the host, but our retrievals return a metallicity consistent with the host within 1$\upsigma$. Our results offer new insights into the formation history of Ross 458c, as well as the efficacy of fitting forward model grids versus retrievals to derive atmospheric properties of directly imaged companions.

\end{abstract}

\keywords{Planet formation --- brown dwarfs --- forward models --- atmospheric retrievals}


\section{Introduction} \label{sec:intro}

\subsection{Background}

Direct imaging surveys have revealed a small population of substellar companions at extremely wide separations from their host stars. These companions can range from “planetary mass” (2-13 $M_J$) to “brown dwarf mass” (13-80 $M_J$), and they orbit their host stars at hundreds to thousands of AU \citep[e.g.][]{Bailey2014, Gauza2015, Zhang2021-C2b}.

A number of theories have been proposed to explain the formation of these companions and their extreme separations. One theory is that they formed through “top-down” gravitational collapse through fragmentation of the molecular cloud that formed the host \citep{Bate2002} \change{or gravitational instability of the protostellar disk \citep{Kratter2016, Adams2025}}. Another theory is that they formed through “bottom-up” core accretion in a protoplanetary disk \citep{Pollack1996}. Molecular cloud fragmentation is consistent with the extreme separations of these companions, but formation through core accretion would require scattering of the companion by a more massive inner body to its current orbital separation. While this is an added difficulty to explaining a core accretion formation scenario, it does not rule it out, as dynamical scattering with other companions could occur \citep[e.g.,][]{Vorobyov2013}.

The IAU discriminator of planets and brown dwarfs is a mass cutoff at the deuterium burning limit (13 $M_J$) \citep{IAU2022}. However, the mass of a companion does not elucidate its formation pathway. A companion could form through core accretion, and subsequently grow to masses higher than 13 $M_J$. Conversely, a companion could form through gravitational collapse of a fragment with mass less than 13 $M_J$, and is only limited by the size at which the fragment is opaque to its own radiation. This limiting mass is called the ``opacity limit," and is typically cited as between 2 and 4 $M_J$ \citep{Whitworth2018}.

One method of determining the formation history of a companion is by analyzing its composition \citep[e.g.][]{Todorov2016,Madhusudhan2019,Zhang2025}. A companion that forms through a “top-down” formation method, e.g. molecular cloud fragmentation or disk instability, would likely have a composition that closely resembles its host star, because the material available to it is the same as the material which formed the host. On the other hand, a companion that forms through core accretion could have a very different composition from its host, because the abundances of molecules in the gas phase change with separation, and they can become enriched in heavy elements due to solid accretion \citep{Oberg2011}. For instance, Jupiter has 3x the solar abundance of carbon, nitrogen, and oxygen \citep{JupiterAbundances}. \change{Outside our solar system, \citet{Thorngren2016} found a trend between mass and metallicity of transiting planets, with low mass planets (which likely formed through core accretion) being metal-enriched.} \change{The directly imaged planet AF Lep b is similarly metal-rich, and at an orbital separation of about 8 AU, the core accretion model is strongly preferred \citep{AFLepb1, AFLepb2, AFLepb3}.} \change{(Metal enrichment has also been found in a few directly imaged systems tens of AU away from their host, such as the exceptional HR 8799 system, but even with core accretion, it is difficult to explain the factor of $\sim$100 higher metal fraction in these objects relative to their hosts; \citealt{HR8799}.)} 

In terms of C/O ratio, \citet{Hoch2023} found that transiting planets inside of 1 AU have a wide scatter, ranging from 0.3 to 1.6, while directly imaged planets outside of tens of AU have C/O ratios ranging from 0.5 to 0.6, very close to the solar value of 0.54 \citep{Asplund2005}. \change{Nearby M stars follow a tight distribution of C/O around the solar value, and in a sample of 46 M stars collected from \citet{Tsuji2014}, \citet{Tsuji2015}, and \citet{Tsuji2016}, all had a C/O ratio between 0.4 and 0.8. Because widely separated companions follow a similar distribution of C/O, they appear to be consistent with a stellar formation pathway at a population level.}

Recent spectroscopic observations from the James Webb Space Telescope (JWST) offer unprecedented constraints on the atmospheres of these companions. \citet{Miles2023} report a direct detection of silicate clouds from their JWST NIRSpec IFU and MIRI MRS spectra of VHS 1256b. \citet{Luhman2023} find that their JWST NIRSpec IFU spectrum of  TWA 27B is not adequately explained by self-consistent cloudless models, and that the discrepancies can likely be accounted for by the presence of clouds, a result supported by a recent analysis by \citet{Zhang2025-TWA27B}. \cite{Faherty2024} detect methane emission features from the Y dwarf W1935, attributing it to possible aurorae. 

Here, we present a JWST NIRSpec Fixed Slit spectrum of Ross 458c, a directly imaged companion at a projected separation of 1100 AU from its binary star host, Ross 458AB, and we use this spectrum to investigate its formation pathway. In Section \ref{sec:ross458}, we describe the properties of the host and companion from the literature. In Section \ref{sec:host_z}, we analyze a spectrum of Ross 458AB to better constrain its metallicity. In Section \ref{sec:data}, we describe the data reduction pipeline and post-reduction processing \change{of our new JWST NIRSpec Fixed Slit spectrum}. In Section \ref{sec:fm}, we use self-consistent forward model grids to fit the spectrum. In Section \ref{sec:retrievals}, we use the \Poseidon\ code to retrieve atmospheric properties of Ross 458c. In Section \ref{sec:discussion}, we discuss our results and their implications on Ross 458c's formation pathway, and we conclude in Section \ref{sec:conclusions}.

\subsection{Ross 458 System} \label{sec:ross458}

\subsubsection{Ross 458AB} \label{sec:host}
Ross 458AB is a M0.5 and M7 binary system at a distance of 11.51$\pm$0.02 pc \citep{Gaia}. \citet{Burgasser2010} \change{placed} age constraints on the system between 150 and 800 Myr. The lower limit \change{was} derived from the absence of Li I absorption and low surface gravity indicators expected in young M dwarfs, whereas the upper limit \change{was} derived from the presence of a prominent H$\alpha$ line and magnetic activity in Ross 458A. \change{It was previously suggested that the Ross 458 system could be a member of the Hyades supercluster, an extended population of stars which includes the 625 Myr Hyades open cluster \citep{Eggen1960, Lebreton1997, Montes2001}, although this was later disputed by \citet{Burgasser2010}. It has also been suggested that it could be a member of the 50 Myr IC 2391 Moving Group \citep{Nakajima2010}, although the IC 2391 would be inconsistent with other estimates of the system's age. Using the BANYAN $\Sigma$ code \citep{Gagne2018}, \citet{Zhang2021b} found a 99.9\% probability that the Ross 458 system is a field system.}

\citet{Burgasser2010} \change{derived} a supersolar metallicity between +0.2 and +0.3 from its position on the $V - K / M_K$ CMD, although they \change{noted} that their results may be biased towards high metallicity due to its young age. \change{Using the procedure from \citet{Mann2013}, \citet{GaidosMann2014} derived a metallicity of 0.25$\pm$0.08 from the H- and K-band spectra, consistent with the results of \citet{Burgasser2010}.}

% Using Equation 2 from \cite{Nissen2013}, we find that these metallicity values correspond to a [C/O] of +0.042 or +0.064 respectively relative to solar. These estimates have a standard deviation of $\pm$0.048 dex. Adopting the solar C/O ratio used by \cite{Nissen2013}, $\rm (C/O)_\odot = 0.58$, this corresponds to a C/O ratio of $0.64\pm0.07$ or $0.67\pm0.07$ respectively.

In Section \ref{sec:host_z}, we re-analyze the host binary to \change{offer new constraints on} its composition.

\subsubsection{Ross 458c} \label{sec:object}

\begin{table*}
    \centering
    \begin{tabular}{l c c c} \hline\hline
         Parameter&  Value&  Units/Notes& Reference\\ \hline
         $\alpha$&  13:00:41.15&  J2000& \cite{Gaia} \\
         $\delta$&  12:21:14.22&  J2000& \cite{Gaia} \\
         Spectral Type&  T8.9$\pm$0.3&  $-$& \cite{Goldman2010}\\
         &  T8&  $-$& \cite{Burgasser2010}\\
         &  T8.5p&  $-$& \cite{Burningham2011}\\
         $T_{\rm eff}$&  $695\pm60$&  K& \cite{Burgasser2010}\\
         &  $650\pm25$&  K& \cite{Burningham2011}\\
         &  $721\pm94$&  K& \cite{Filippazzo2015}\\
         &  $804^{+30}_{-29}$ &  K& \cite{Zhang2021a}\\
 & $762.64^{+6.85}_{-1.81}$& K&\cite{Zalesky2022}\\
         &  $722^{+11}_{-12}$&  K& \cite{Gaarn2023}\\
         % & $651.2\pm0.3$& K, ExoREM&This work\\
 & $771.1^{+7.2}_{-8.0}$& K, \Poseidon&This work\\
         $\log g$ (cgs)&  $4$&  $-$& \cite{Burgasser2010}\\
         &  $4-4.7$ &  $-$& \cite{Burningham2011}\\
         &  $3.7-4$& $-$& \cite{Morley2012}\\
 & $3.74^{+0.33}_{-0.17}$& $-$&\cite{Zalesky2022}\\
         &  $4.50\pm0.07$ & $-$& \cite{Gaarn2023}\\
 % & $4.42$& ExoREM&This work\\
 & $4.12^{+0.04}_{-0.04}$& \Poseidon&This work\\
         $\log_{10} L_{bol}/\rm L_\odot$&  $-5.62\pm0.03$ & $-$& \cite{Burgasser2010} \\
         &  $-5.61$&  $-$& \cite{Burningham2011}\\
         &  $-5.27\pm0.03$&  $-$& \cite{Gaarn2023}\\
         Metallicity&  $-0.06\pm0.20$&  $-$& \cite{Goldman2010}\\
         &  $0$&  Assumed& \cite{Burgasser2010}\\
         &  $0$&  Assumed& \cite{Burningham2011}\\
 & $-0.20^{+0.14}_{-0.09}$& [M/H]&\cite{Zalesky2022}\\
         &  $0.18\pm0.04$&  [C/H]& \cite{Gaarn2023}\\
 % & $0.25\pm0.01$& ExoREM&This work\\
 & $0.05^{+0.03}_{-0.02}$& [M/H], \Poseidon&This work\\
         C/O& $1.97^{+0.13}_{-0.14}$& CH$_4$/H$_2$O&\cite{Gaarn2023}\\
 & $0.51^{+0.12}_{-0.07}$& $-$&\cite{Zalesky2022}\\
         % & $0.64$& ExoREM&This work\\
 & $0.68^{+0.03}_{-0.02}$& \Poseidon&This work\\
         Age &  $0.4-0.8$&  Gyr& \cite{West2008}\\
         &  $0.15-0.8$&  Gyr& \cite{Burgasser2010}\\
         &  $<1$&  Gyr& \cite{Manjavacas2019}\\
         Mass&  $5-14$&  $M_J$& \cite{Goldman2010}\\
         &  $6.29-11.52$&  $M_J$& \cite{Burgasser2010}\\
         &  $5-20$&  $M_J$& \cite{Burningham2011}\\
         &  $2.3^{+2.3}_{-1.2}$&  $M_J$& \cite{Zhang2021a}\\
         &  $27^{+4}_{-4}$&  $M_J$& \cite{Gaarn2023}\\
 % & $16.2\pm0.1$& $M_J$, ExoREM&This work\\
 & $8.24^{+0.80}_{-0.75}$& $M_J$, \Poseidon&This work\\
         Radius&  $1.19-1.29$&  $R_J$& \cite{Burgasser2010}\\
         &  $1.01-1.23$&  $R_J$& \cite{Burningham2011}\\
         &  $0.68\pm0.06$&  $R_J$& \cite{Zhang2021a}\\
 & $0.87^{+0.09}_{-0.05}$& $R_J$&\cite{Zalesky2022}\\
         &  $1.45\pm0.06$&  $R_J$& \cite{Gaarn2023}\\
 % & $1.30$& $R_J$, ExoREM&This work\\
 & $0.85\pm0.02$& $R_J$, \Poseidon&This work\\ \hline
    \end{tabular}
    \caption{Literature values for Ross 458c parameters. The table was originally compiled by \cite{Gaarn2023} and extended by this work. We report values from ExoREM, our preferred model among the forward model grids, and from the \Poseidon\ model with Na$_2$S clouds.}
    \label{tab:literature}
\end{table*}

Ross 458c is a T8 companion to Ross 458AB, \change{discovered concurrently by \citet{Goldman2010} and \citet{Scholz2010}}, at an angular separation of 102$''$. This gives the companion a projected separation of 1168 AU from the primary. Existing literature values for the parameters of Ross 458c are listed in Table \ref{tab:literature}.

Previous studies have used \change{spectra} from the FIRE spectrograph on Magellan \change{and the IRTF/SpeX spectrograph}, and have found effective temperatures between 650 and 800 K. \citet{Burgasser2010} found 650$\pm$25 K from fitting self-consistent forward models \change{to the FIRE spectrum}. \citet{Burningham2011} found 695$\pm$60 K, and \citet{Filippazzo2015} found 721$\pm$94 K using the bolometric luminosity. \citet{Zhang2021a} found $804^{+30}_{-29}$ K from \change{fitting forward models to the IRTF/SpeX spectrum, and \citet{Zalesky2022} found $762.64^{+6.85}_{-1.81}$ K from a retrieval on the IRTF/SpeX spectrum.} Finally, \citet{Gaarn2023} found $722^{+11}_{-12}$ K using the \Brewster\ retrieval code \change{applied to the FIRE spectrum}.

\citet{Burgasser2010} found that Ross 458c is better fit by a cloudy model, a result which was surprising at the time, as clouds were thought to sink below the photosphere in T dwarfs. \citet{Burningham2011} corroborated this finding, although they found that thinner clouds from the BT Settl models are preferred over the Saumon and Marley models used in \citet{Burgasser2010}. Similarly, \citet{Gaarn2023} found that virtually any cloudy model is preferred over a cloud-free model in their retrieval analysis of Ross 458c. %\note{Their preferred model was a power-law opacity slab cloud, followed by MgSiO$_3$ clouds and Na$_2$S clouds.} 
On a population level, \citet{Zhang2021b} found that the cloud-free and chemical equilibrium assumptions used by the Sonora Bobcat forward models ``do not adequately reproduce late T-dwarf spectra.''

While \citet{Gaarn2023} corroborated the presence of clouds on Ross 458c, their most surprising result was that their preferred model suggests a C/O ratio of $1.97^{+0.13}_{-0.14}$, too high to be consistent with its nearby M dwarf host, which would be unlikely to be higher than 0.8 \citep[][]{Nakajima2016}. (They derive their C/O ratio from the CH$_4$/H$_2$O ratio, because they do not detect CO or CO$_2$ in the NIR spectrum.) They explored how the C/O ratio might change due to a variety of factors: telluric contamination, oxygen condensing into clouds, and missing CO and CO$_2$. None of these factors were able to explain Ross 458c's high C/O ratio. They interpreted this as suggesting that Ross 458c likely formed through a planetary formation pathway. In Sections \ref{sec:fm} and \ref{sec:retrievals}, we investigate whether this high C/O ratio holds with our JWST NIRSpec spectrum.

\section{Ross 458AB re-analysis}\label{sec:host_z}
Before we analyze Ross 458c, it is worth revisiting its host stars, an M0.5+M7 binary system. We intend to compare the composition of the wide-orbit companion to that of its host, \change{and to do this, we first need to constrain the composition of the primary.}% but the composition of the host is not well constrained.} % \citet{Burgasser2010} estimates its metallicity to be +0.2 or +0.3, depending on calibration differences, with both measurements coming from its placement on a CMD. They note that their estimate may be biased toward high metallicity, but 
%Little work has been done in the years following the analysis of \citet{Burgasser2010} to further constrain the metallicity of Ross 458AB.

\change{In this section, we compare an archival SpeX spectrum of Ross 458AB to interpolated grids of PHOENIX stellar atmosphere models to constrain its metallicity \citep{PHOENIX}.}

% In this section, we apply the same forward model analysis that we apply to the companion in Section \ref{sec:fm} to constrain the metallicity of the host. We use an archival SpeX spectrum of Ross 458AB, and we compare it to interpolated grids of PHOENIX stellar atmosphere models to constrain its metallicity \citep{PHOENIX}.

\subsection{Data}

The IRTF/SpeX \citep{SpeX} data of the host stars Ross 458AB were downloaded from the IRTF archive\footnote{\url{https://irsa.ipac.caltech.edu/applications/irtf/}} (PI: Z. Zhang). The observations were carried out with an average airmass of 1.52 and seeing of 0.5-0.6", using the ShortXD grating and 0.3$\times$15" slit, covering a wavelength range of 0.70$-$2.55 $\upmu$m \change{at a resolving power of $\sim$2000}. A total of eight spectra were obtained employing an ABBAABBA dither pattern, with individual exposures of 29.65s each for a total of 237.2s integration time. The raw data frames were reduced with the Xspextool\footnote{\url{https://irtfweb.ifa.hawaii.edu/~spex/observer/}} pipeline, correcting for flatfielding, tracing, wavelength- and flux calibration \citep{Spextool}. Telluric absorption was corrected for utilizing the standard A0 star HD 114381. The eight individual spectra were then scaled to the median and sequentially combined to a single, reduced spectrum. The errors were derived by taking the standard error around the median.

\subsection{Methodology}

For our analysis of Ross 458AB, we use the PHOENIX stellar atmosphere models developed by \citet{PHOENIX}. These models vary in $T_{\rm eff}$, $\log g$, [Fe/H], and [$\alpha$/Fe], although [$\alpha$/Fe] is only used in the low metallicity models, which we do not consider in our analysis. The abundances of elements in the PHOENIX models are scaled from solar abundances from \citet{Asplund2009}, so the C/O ratio for each of the models is assumed to be solar \citep{PHOENIX}. In our wavelength range of interest, 0.7$-$2.5 \textmu m, the PHOENIX models are sampled at a resolving power of 500,000.

From \citet{Burgasser2010}, Ross 458AB are an M0.5+M7 pair. By plotting approximate PHOENIX models with appropriate temperature, radius and mass from \citet{MSparams}, we find that A is approximately 100 times brighter than B at all wavelengths, which is comparable to the signal-to-noise of the spectrum. After a series of tests where we attempted to fit the spectrum with A alone, or with A and B, we found that considering the model uncertainty and interpolation uncertainty makes the contribution from B less than the total uncertainty in the fit. These tests are described in more detail in Appendix \ref{app:joint}. Because of this, we decided to only consider the contribution from A in the fit.

The PHOENIX models described in \citet{PHOENIX} range from $2,300 \le T_{\rm eff} \le 12,000$, $0.0 \le \log g \le 6.0$, and $-4.0 \le \rm [Fe/H] \le +1.0$, but we restrict our analysis to the range shown in Table \ref{tab:fm_grids}, which is derived from \change{values for main sequence stars in} \citet{MSparams}. \change{(We verified that Ross 458A is on the main sequence using the AMES-Cond isochrones, from which we found that a 0.54 $M_\odot$ star would reach the main sequence after about 90 Myr; \citealt{Allard2001, Baraffe2003}. The lower limit on the age of the Ross 458 system is 150 Myr, so it should have reached the main sequence.)} Because Ross 458A is an M0.5V star, we allow our priors to range between the listed values for K9V through M1.5V. We note that although PHOENIX varies in $\log g$, we place a prior on the mass, which is calculated from $\log g$ and radius, because mass is listed in \citet{MSparams}.

We convolve the PHOENIX models to the same resolving power and wavelength grid as the SpeX spectrum of Ross 458AB, and we use the \texttt{RegularGridInterpolator} function in the \texttt{scipy} package to interpolate these convolved forward models \citep{scipy, RegularGridInterpolator}. We then use the minimize function from \texttt{scipy.optimize} to find the maximum log-likelihood of our interpolated models with our data, using the values for an M0.5V star from \citet{MSparams} as our initial parameters. Finally, we use the parameters which maximize log-likelihood as our initial condition for a Markov chain Monte Carlo (MCMC) run using the \texttt{emcee} Python package \citep{emcee}.

Previous studies have used a wavelength-independent parameter to characterize unknown uncertainties that may affect either the data or the model \cite{Line2015}. \cite{Piette2020} found that there are uncertainties in calculating forward models that scale with the flux, so they introduce a wavelength-dependent parameter which increases errors as a percentage of the flux in a given wavelength bin.

Here, we include both forms of error inflation parameters to model the systematic uncertainties, labeled $\log x_{tol}$ and $b$. We use $\log x_{tol}$, the (base-10) logarithm of $x_{tol}$, to parameterize uncertainties which scale with the flux, and we use $b$ to parameterize uncertainties that are constant in wavelength. These parameters also inform us about the quality of the model grid with respect to the data. A higher $\log x_{tol}$ or $b$ implies a worse fit.

The log-likelihood function used by the minimize function and the \texttt{emcee} script is given by Equation \ref{eq:likelihood}, where $F_\lambda$ is the flux density from the data, $M$ is the flux density from the model, and $s^2$ is given by Equation \ref{eq:error}, where $\sigma$ is the uncertainty in the data, and $x_{tol}$ and $b$ are described above.

\begin{equation}
    \label{eq:likelihood}
    \ln\mathcal{L} = -\frac{1}{2}\sum_\lambda{\left[\left(\frac{F_\lambda(\lambda) - M(\lambda)}{s(\lambda)}\right)^2 + \ln(2\pi s^2)\right]}
\end{equation}

\begin{equation}
    \label{eq:error}
    s^2 = \sigma^2 + (x_{tol}\times F_\lambda)^2 + 10^b
\end{equation}

\subsection{Results}
\begin{figure*}
    \centering
    \includegraphics[width=1\linewidth]{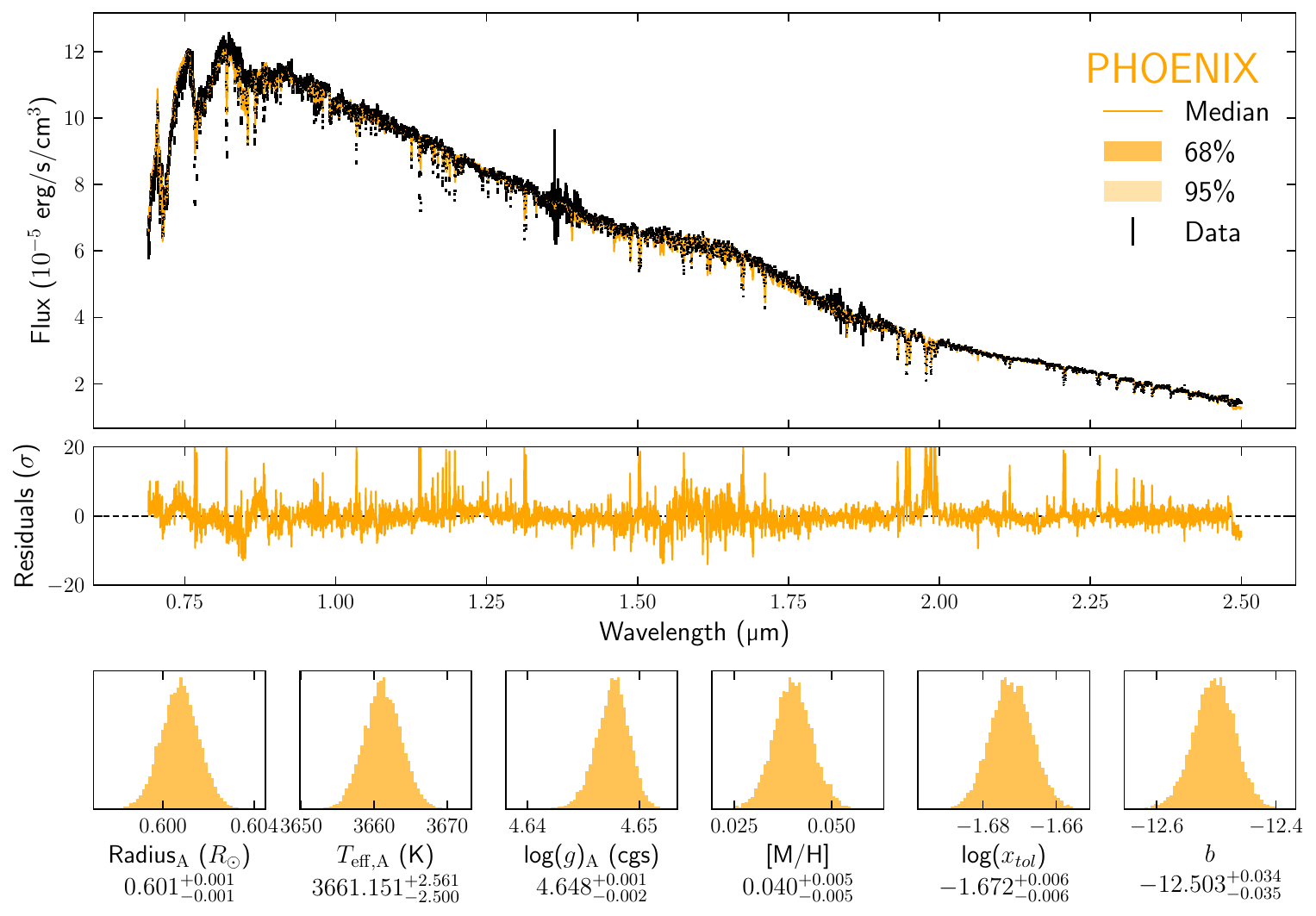}
    \caption{Summary figure of the fit to Ross 458A using interpolated grids of PHOENIX models. Top panel: the median, $1\upsigma$, and $2\upsigma$ distributions of a sample of one hundred spectra randomly sampled from the resulting parameter distribution, compared to the data. The $1\upsigma$ and $2\upsigma$ distributions are smaller than the width of the median line. Middle panel: the difference between the model and the data flux divided by the uncertainty in the data \change{without considering model uncertainties parameterized by $x_{tol}$ and $b$}. Bottom panels: probability density functions of each of the freely varying parameters.}
    \label{fig:phoenix}
\end{figure*}

The results of our $\texttt{emcee}$ run are given in Figure \ref{fig:phoenix}, with the resulting probability distributions of the free parameters shown in the bottom panels.

% \begin{table*}
%     \centering
%     \begin{tabular}{|c|c|c|c|c|c|c|} \hline
%         Model Grid & Radius ($R_\odot$)& $T_{\rm eff}$ (K) & $\log g$ (cgs)& [Fe/H] & $\log x_{tol}$&$b$\\ \hline
%         PHOENIX& $0.601^{+0.001}_{-0.001}$& $3661.15^{+2.56}_{-2.50}$& $4.648^{+0.001}_{-0.002}$& $+0.040^{+0.005}_{-0.005}$ & $-1.672^{+0.006}_{-0.006}$&$-6.252^{+0.017}_{-0.018}$\\\hline
%     \end{tabular}
%     \caption{PHOENIX results table}
%     \label{tab:host_results}
% \end{table*}

Using the error inflation terms to examine the quality of the fit, we find that $x_{tol}$ corresponds to an additional variance of about 2\% of the flux, while $b$ corresponds to \change{an additional uncertainty of $5.598\times10^{-7}\rm\ erg/s/cm^3$, comparable to the median uncertainty in the data, $5.087\times10^{-7}\rm\ erg/s/cm^3$, added in quadrature}. From this, we conclude that the $\texttt{emcee}$ chain did not need to increase the errors by much to obtain a relatively good fit. We find this to be a decent fit, given the inherent uncertainty in the models and interpolation process.

\change{For an M0.5V star, we expect the radius to be about 0.54 $R_\odot$ and the effective temperature to be about 3770 $\rm K$ \citep{MSparams}.} We find the model radius to be higher than expected, while the effective temperature is lower than expected. However, both values are well constrained inside the parameter space.

\begin{figure}
    \centering
    \includegraphics[width=\linewidth]{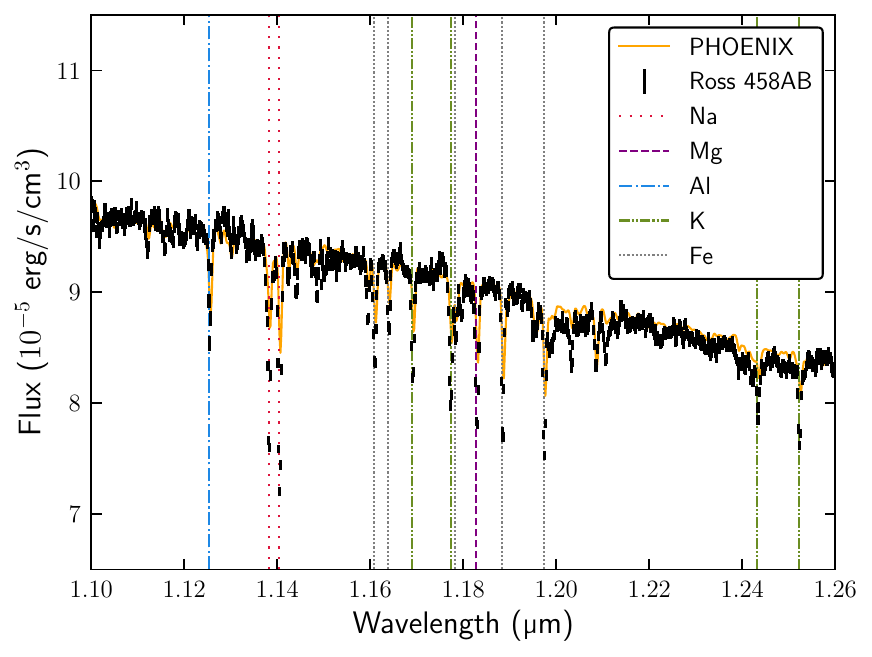}
    \caption{\change{Median PHOENIX model compared to the spectrum of Ross 458AB between 1.10 and 1.26 $\upmu$m, with various atomic lines from NIST labeled \citep{NIST}. The PHOENIX model does not go as deep as the data at many of the atomic lines, leading to spikes in the residuals.}}
    \label{fig:atomics}
\end{figure}

We find a good match between the median spectrum and the data in the continuum and molecular features, but we find that \change{it} has difficulty fitting some of the atomic lines. \change{For illustration, we show in Figure \ref{fig:atomics} a zoomed-in view between 1.10 - 1.26 $\upmu$m, with several atomic lines marked.} In general, the model atomic lines do not go as deep as the data, indicating a possible underfitting of $\log g$.% \textcolor{magenta}{It especially misses some lines from 1.9 to 2.1 microns, and I've been unable to figure out which lines these are. Should I just end the paragraph and move on from it?}

We find that the metallicity derived from this analysis is lower than both previous estimates from \citet{Burgasser2010} \change{and the value from \citet{GaidosMann2014}} at $+0.040^{+0.005}_{-0.005}$. This is consistent with the prediction from \citet{Burgasser2010} that their derived metallicity may be an overestimate due to the system's young age, \change{but \citet{GaidosMann2014} make no such qualification.} %We find that their metallicity was likely an overestimate, and that Ross 458AB is in fact much closer to solar metallicity. 
Because the PHOENIX models assume solar C/O, we are unable to constrain the C/O of the primary using this analysis.%, but because the metallicity is near solar, we assume the C/O ratio of the primary is also approximately solar. %Using the relation from \citet{Nissen2013}, we derive a C/O ratio of $0.59\pm0.07$. This finding that Ross 458AB has a composition very close to solar has important consequences for our interpretation of Ross 458c.

\section{Ross 458c Data} \label{sec:data}

\subsection{Observations and Reduction} \label{sec:observation}

The JWST/NIRSpec Fixed Slit observations of Ross 458c were part of JWST GTO program \#1292 (PI: J. Lunine) using the F070LP, F100LP, and F170LP filters and the ALLSLITS subarray. The observations were carried out on 2023-07-04, starting at UTC 04:39:09.608 and ending at UTC 10:59:39.239. The F070LP and F100LP filters used the G140H grating, and the F170LP filter used the G235H grating. Each observation with a given filter/grating used two dither positions, two slits (S200A1 and S200A2), and a NRSRAPID readout pattern. The exposure time per dither per slit for each of the filter choices was the following: 17 minutes for F070LP, 39 minutes for F100LP, and 30 minutes for F170LP, for a total exposure time of about 5 hours and 10 minutes. The total wavelength range covers 0.8$-$3.1 \textmu m at a resolving power of $\sim$2700. The raw {\it JWST} data used in this paper can be found in MAST: \dataset[10.17909/70vg-1v90]{http://dx.doi.org/10.17909/70vg-1v90}.

The data were reduced using the standard JWST pipeline version 1.14.0. The CRDS context used by the JWST pipeline was \texttt{jwst\_1230.pmap}.

All frames are run through Stage 1, which performs detector-level corrections and converts detector ramps into slope images. They are then processed through Stage 2, which performs more instrument corrections and produces calibrated slope images. Image-from-image background subtraction was performed in Stage 2, using the background from one dither to subtract from the equivalent dither, rather than Master Background Subtraction in Stage 3. Finally, they are processed through Stage 3, which combines the calibrated slope images into a one-dimensional extracted spectrum. This process was carried out separately for the three filters, producing three extracted spectra. 

\begin{figure*}
    \centering
    \includegraphics[width=1\linewidth]{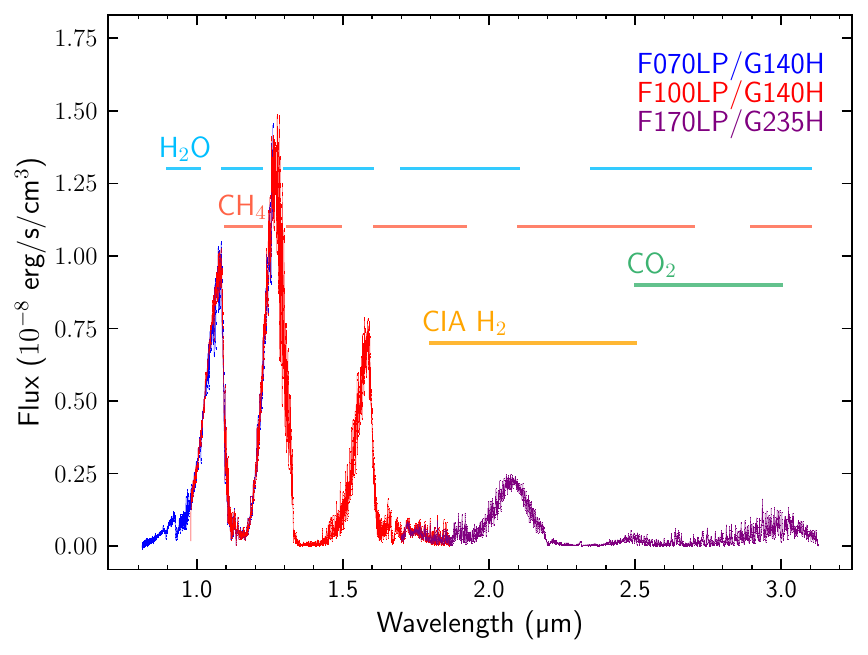}
    \caption{JWST NIRSpec Fixed Slit spectra of Ross 458c using F070LP/G140H, F100LP/G140H, and F170LP/G235H filter/grating combinations. The F100LP/G140H flux has been calibrated to UKIDSS J-band photometry, and the F070LP/G140H and F170LP/G235H flux have been calibrated to the F100LP/G140H flux by equalizing the mean of the overlapping regions. The spectra range from 0.8 to 3.1 \textmu m at an average resolving power of R$\sim$2700. Locations of prominent absorption bands (H$_2$O, CH$_4$, and collisionally induced H$_2$) are labeled.}
    \label{fig:spectra}
\end{figure*}

We calibrated the F100LP/G140H spectrum to UKIDSS J-band photometry used in \citet{Burningham2011} and \citet{Gaarn2023}. We then calibrated the F070LP/G140H and F170LP/G235H spectra to the F100LP/G140H spectrum by applying a multiplicative factor that equalizes the mean of the regions of overlap. The F070LP/G140H, F100LP/G140H, and F170LP/G235H flux densities are multiplied by factors of 0.965, 0.948, and 0.981 respectively. The reduced spectra of Ross 458c are shown in Figure \ref{fig:spectra}.

\subsection{Error adjustment} \label{sec:error}

% We investigated the uncertainties returned by the pipeline, and found that they were underestimated when compared to the statistical uncertainties of the individual spectra. Previous spectral modeling efforts for directly imaged spectra tend to include an "error inflation" parameter in the model, often by as much as a factor of [x], e.g. [Source]. Rather than inflate the errors arbitrarily, we performed our own analysis to determine an appropriate error inflation factor for each of the spectra based on empirically calculated uncertainties. \textcolor{magenta}{Going to change this paragraph. Ryan explained to me how an error inflation parameter is still useful to tell you how poor the model is, so I think we'll still include it in our retrievals. Need to motivate the pre-inflation outside the context of retrievals.}

We investigated the uncertainties returned by the JWST pipeline, and found that they were underestimated when compared to the statistical uncertainties of the individual spectra.

At each wavelength bin, we calculated the statistical uncertainty of the individual spectra. In general, there were four independent samples per wavelength bin (2 slits, 2 dither positions), but in the detector gaps, there were only two independent samples (1 slit, 2 dither positions). We multiplied the statistical uncertainty by $\sqrt{\frac{N-1}{N}}$, where $N$ is the number of samples in the wavelength bin, to account for the small sample size, following the method outlined in \citet{StatsTechniques}.

\begin{figure}
    \centering
    \includegraphics[width=\linewidth]{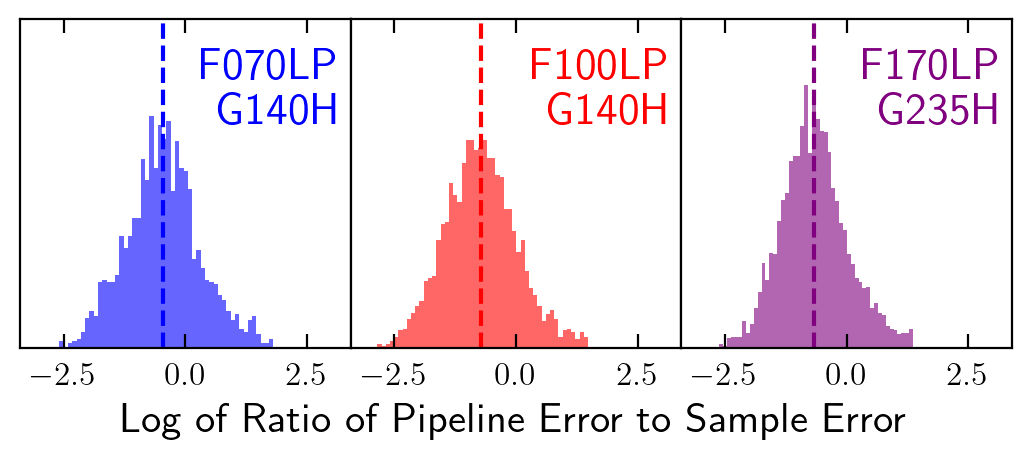}
    \caption{Histogram of the ratio of the error returned by the JWST pipeline to the empirical error measured by comparing the individual frames at a given pixel value, plotted in log space. Each filter/grating combination peaks at a negative value, indicating the pipeline is underestimating the errors compared to the spread of the indiviudal samples. To correct this, we increase the errors by a factor of 1.57 for F070LP/G140H, 2.06 for F100LP/G140H, and 1.94 for F170LP/G235H.}
    \label{fig:offset}
\end{figure}

We calculated the ratio of the error returned by the pipeline to the statistical uncertainty of the samples, took the log of these ratios, and performed an iterative sigma clip at $3\upsigma$. We plotted the log of these ratios in histograms for each filter. These histograms are shown in Figure \ref{fig:offset}.

We expect the histograms to be centered at 0, indicating agreement between the pipeline error and the statistical uncertainty. Because they are each centered at negative values, we infer that the pipeline error is, in general, less than the statistical uncertainty of the samples. To account for this, we increase the errors by a constant factor of 1.57 for F070LP/G140H, 2.06 for F100LP/G140H, and 1.94 for F170LP/G235H. \change{After adjusting the errors, the average S/N of the resulting spectra is $\sim$60.}

\section{Forward Models} \label{sec:fm}

Brown dwarf atmospheres are often studied by comparing to grids of forward models \citep[e.g.][]{Burgasser2010, Hoch2024, Tu2024, COCONUTS-2b}. These grids contain precomputed theoretical model spectra which rely on a set of assumptions regarding input physics and chemistry (i.e. equilibrium chemistry, vertical mixing, presence/absence of clouds), and vary in effective temperature, gravity, metallicity, and other parameters \citep[e.g.][]{Morley2012, Linder2019, ATMO2020}. Traditionally, these comparisons are a simple least-squares analysis which allows one to derive a best-fitting set of parameters. However, this approach is limited by the often coarse sampling of parameter space in forward model grids. For our analysis, we develop an interpolator to probe the space between these grid points, equivalent to our work in Section \ref{sec:host_z}. We describe our forward model grids in Section \ref{sec:fm_grids}, and we describe our interpolation and fitting process in Section \ref{sec:fm_methods}.

\subsection{Model Grids} \label{sec:fm_grids}

\subsubsection{Sonora Bobcat}
The Sonora Bobcat grid of forward models assumes a cloudless atmosphere in chemical equilibrium \citep{Bobcat}. These model spectra are made to be self-consistent with evolution models, which include assumptions about convection and equations of state. \citet{Zhang2021b} find that these models do not fit late T-dwarf spectra, but they are a useful starting point for our forward model analysis due to the simplicity of the parameter space.

The Sonora Bobcat grid ranges from $200 \le T_{\rm eff} \le 2400\rm\ K$, $2.5 \le \log g \le 5.5$, and $-0.5 \le [\rm M/H] \le +0.5$. For the set of spectra that vary in metallicity, C/O is fixed to the solar value from \citet{Lodders2010}, 0.458. We restrict our analysis to the grid range shown in Table \ref{tab:fm_grids}, which encompasses a wide range of $\log g$ and all previously derived effective temperatures for Ross 458c.

\subsubsection{Sonora Elf Owl}
The Sonora Elf Owl forward models assume a cloudless atmosphere, but they allow for disequilibrium chemistry due to vertical mixing \citep{Mukherjee2024}.  Disequilibrium chemistry has been found to be important in brown dwarf atmospheres \citep[e.g.][]{COCONUTS-2b}. It is typically assumed that vertical mixing can be modeled as a diffusion process, so \citet{Mukherjee2024} adopt a vertical eddy diffusion parameter $K_{zz}$ to parameterize the vertical mixing. The Sonora Elf Owl models allow the diffusion parameter to vary from $2 \le \log_{10}\left(K_{zz} [\rm cm^2/s]\right)\le 9$.

Sonora Elf Owl spans a similar range in $T_{\rm eff}$ and $\log g$ to Sonora Bobcat -- $275 \le T_{\rm eff} \le 2400\rm\ K$ and $3.25 \le \log g \le 5.5$ -- but it allows metallicity to range from $-1.0 \le [\rm M/H] \le +1.0$. It also allows for a variable C/O ratio, from $0.22 \le \rm C/O \le 1.14$. The grid range used for our analysis is shown in Table \ref{tab:fm_grids}.

\subsubsection{ExoREM}
ExoREM is a radiative-convective equilibrium model for young giant companions on the planet-brown dwarf boundary, developed by \citet{Baudino2015}. ExoREM solves for radiative-convective equilibrium, assuming that the net flux is conservative, and neglecting stellar heating. The original ExoREM models assumed a cloudless atmosphere in chemical equilibrium, but \change{the public} ExoREM models provided by \citet{Charnay2018} \change{allow for iron and silicate clouds}. The cloud parameterization uses simple microphysics, and the models allow for disequilibrium chemistry due to vertical mixing. These are the models that we use for our analysis, and a full description of these models can be found in \citet{Baudino2015} and \citet{Charnay2018}.

The publicly available R=20000 ExoREM models range from $400 \le T_{\rm eff} \le 2000\rm\ K$, $3.0 \le \log g \le 5.0$, $-0.5 \le [\rm M/H] \le +2.0$, and $0.10 \le \rm C/O \le 0.80$, but we restrict our analysis to the range shown in Table \ref{tab:fm_grids}.

\subsection{Methodology} \label{sec:fm_methods}

To begin, we develop interpolators using the \texttt{RegularGridInterpolator} function in the \texttt{scipy} package to probe the parameter space between the grid points of our forward model grids. \texttt{RegularGridInterpolator} is the optimal interpolator to use for regularly spaced grids, but it breaks down if any of the grid points are missing. Sonora Bobcat and ExoREM in particular are incomplete grids, so we fill in the missing grid points by interpolating from the two nearest models in $\log g$ space. For the Sonora Bobcat grid, this fills in each of the missing models, but for ExoREM, there are two grid points in our range that are unable to be filled in with this method. For these, we interpolate between the two nearest models in C/O space.

We note that any interpolation process introduces an additional uncertainty that is not accounted for in the posterior, and so the uncertainties derived from this method should be considered as a lower limit to the parameter uncertainties.

We start by performing a simple least-squares fit using each of the publicly available models in each of the grids, with radius ranging from $0.5 - 5.0\  R_J$ in increments of $0.1\ R_J$. We then use the minimize function from \texttt{scipy.optimize} to find the maximum log-likelihood of our interpolated models with our data, using the best-fitting model from the least-squares fit as our initial parameters. Finally, we use the parameters which maximize log-likelihood as our initial condition for an \texttt{emcee} run.

We include the same two error inflation parameters, $\log x_{tol}$ and $b$, that we used to fit the spectrum of Ross 458AB in Section \ref{sec:host_z}. Our likelihood function is described in the same way, using Equations \ref{eq:likelihood} and \ref{eq:error}.

\begin{table*}
    \centering
    \begin{tabular}{c|c c c c c c c c} \hline\hline
         Object&Model Grid & Radius& $T_{\rm eff}$ (K) &  Mass&$\log g\rm\ (cgs)$ & [M/H] & C/O  & $\log K_{zz}\rm\ (cgs)$\\ \hline 
  Ross 458A&PHOENIX& (0.48, 0.61) $R_\odot$& (3620, 3930)&  (0.47, 0.59) $M_\odot$&$-$& (0.0, +1.0)& 0.55 (fixed)&$-$\\ \hline
         Ross 458c&Sonora Bobcat & (0.5, 5.0) $R_J$& (500, 1000) &  $-$&(3.0, 5.5) & (-0.5, +0.5) & 0.458 (fixed) & $-$ \\
         &Sonora Elf Owl& (0.5, 5.0) $R_J$& (575, 900) &  $-$ &(3.5, 5.0) & (-0.5, +1.0) & (0.22, 1.14) & (2, 9) \\
         &ExoREM & (0.5, 5.0) $R_J$& (500, 800) &  $-$&(3.0, 5.0) & (-0.5, +1.5) & (0.40, 0.80) & $-$\\ \hline
    \end{tabular}
    \caption{Prior range explored for each of the forward model grids used to fit Ross 458A and Ross 458c. \change{The prior range for Ross 458c is wide enough in $T_{\rm eff}$ to cover nearly all previously derived $T_{\rm eff}$ values and a wide range of each of the other variables. We check our prior range choices by examining the resulting parameter distributions, where we find that none of them approach the edge of the priors.}}
    \label{tab:fm_grids}
\end{table*}

\begin{figure*}
    \centering
    \includegraphics[width=1\linewidth]{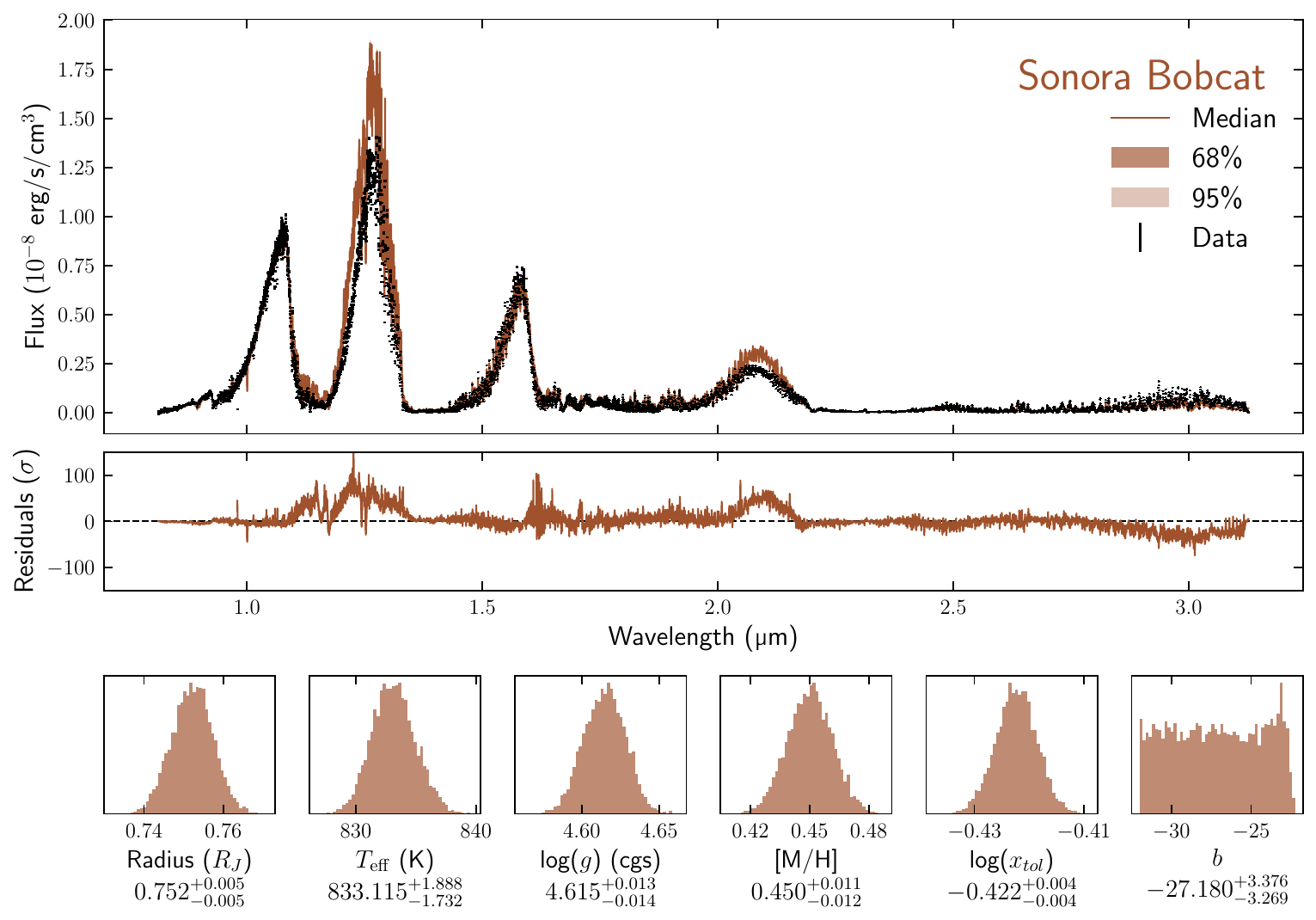}
    \caption{Summary figure of the fit to Ross 458c using interpolated grids of Sonora Bobcat models. The description of each panel is the same as in Figure \ref{fig:phoenix}.}
    \label{fig:bobcat}
\end{figure*}

\begin{figure*}
    \centering
    \includegraphics[width=1\linewidth]{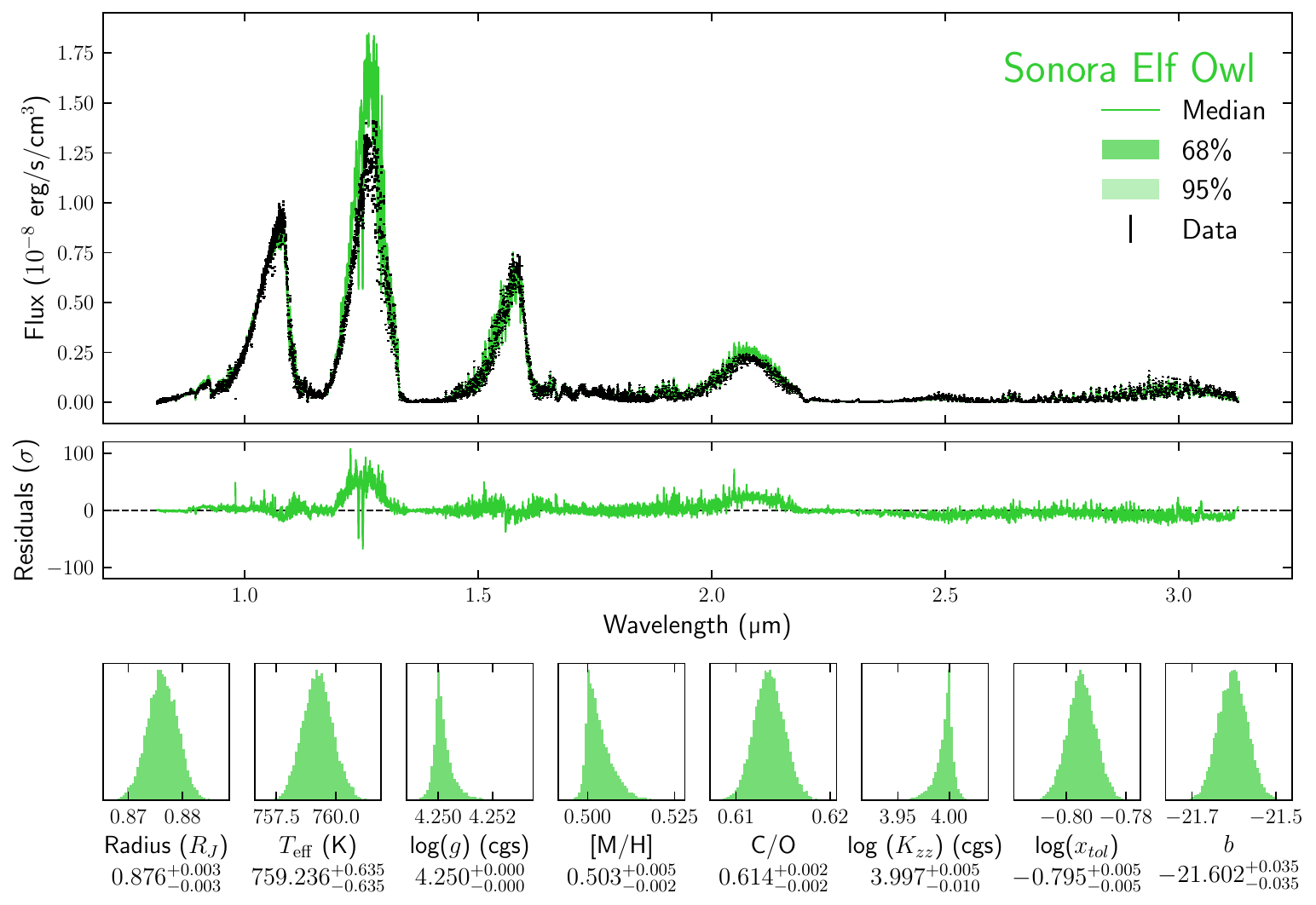}
    \caption{Summary figure of the fit to Ross 458c using interpolated grids of Sonora Elf Owl models. The description of each panel is the same as in Figure \ref{fig:phoenix}.}
    \label{fig:elfowl}
\end{figure*}

\begin{figure*}
    \centering
    \includegraphics[width=1\linewidth]{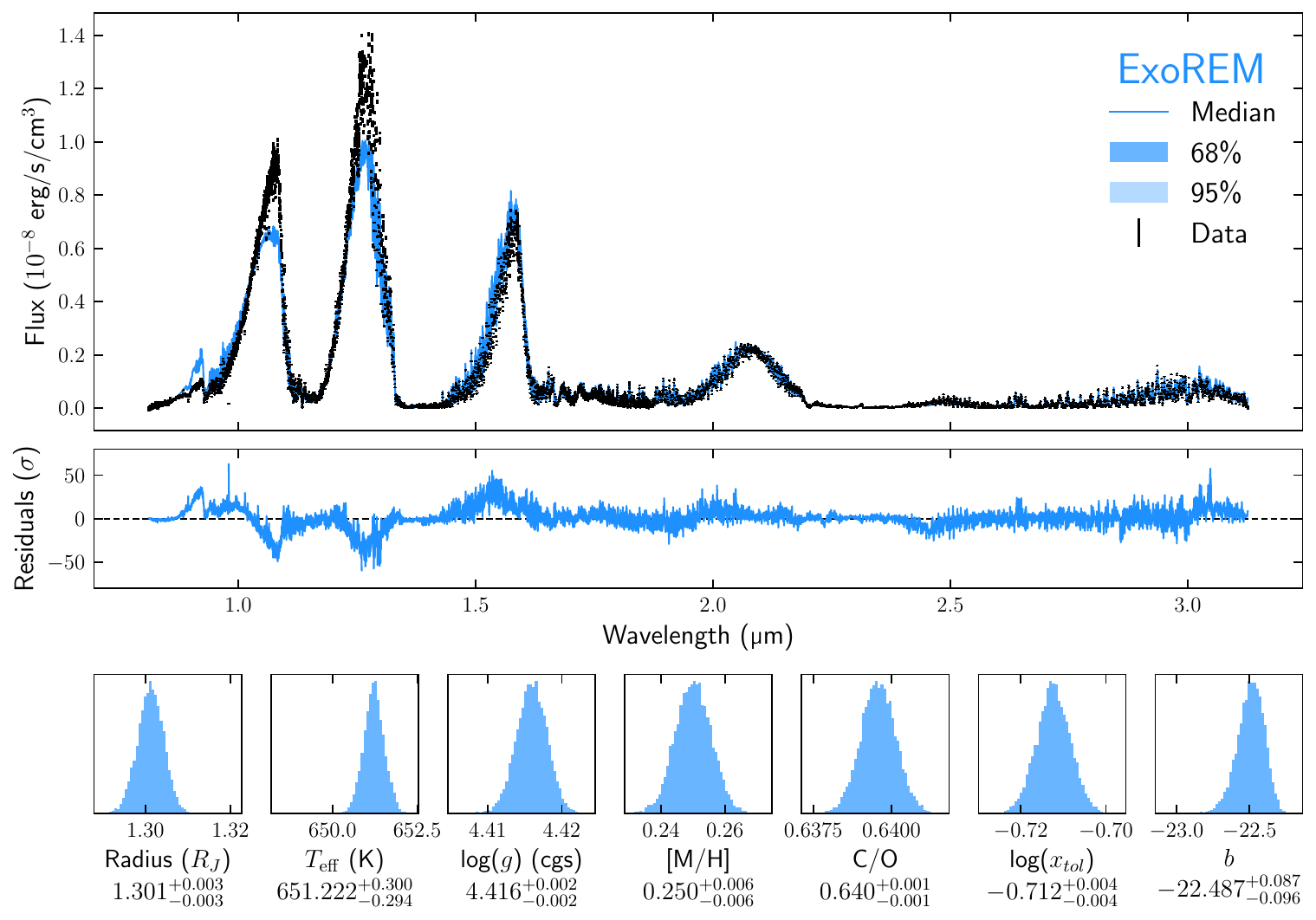}
    \caption{Summary figure of the fit to Ross 458c using interpolated grids of ExoREM models. The description of each panel is the same as in Figure \ref{fig:phoenix}.}
    \label{fig:exorem}
\end{figure*}

\subsection{Results} \label{sec:results}

We present our results in Figures \ref{fig:bobcat}-\ref{fig:exorem}. The resulting parameters are presented in histograms in the bottom panels and summarized in Table \ref{tab:results}.

From the results, we find that the uncertainties on the atmospheric parameters are far too small to be physically reasonable. This is expected, as the interpolation process introduces additional uncertainties that are not accounted for in the posterior distribution.

Visually, we see that Sonora Bobcat provides a reasonable fit to the Y- and H- bands, but greatly overestimates the J- and K- bands, and underestimates the molecular features around 3.0 $\upmu$m. Sonora Elf Owl provides a similar fit, except with less discrepancy in the K-band and at 3.0 $\upmu$m. ExoREM, on the other hand, underestimates the Y- and J- bands, slightly overestimates the H-band, and provides a reasonable fit to the remainder of the spectrum.

We can also compare the quality of the fits by comparing the $\log x_{tol}$ and $b$ parameters \textemdash a higher $\log x_{tol}$ and $b$ indicates that the model had to increase the uncertainties more to maximize the likelihood function. We find that the likelihood of the Sonora Bobcat models did not improve with an increase of $b$, but it did prefer an exceptionally high $\log x_{tol}$, corresponding to about 38\% of the flux added in quadrature to the nominal errors. This is the poorest fit of the three models, consistent with the findings of \citet{Zhang2021b}, who found that the Sonora Bobcat models do not fit well to late T-dwarfs, suggesting that complex atmospheric processes are required to explain these objects. The Sonora Elf Owl and ExoREM fits, on the other hand, constrained a best-fitting $b$ parameter, and also had much lower $\log x_{tol}$, corresponding to 16\% and 19\% of the flux respectively. While the Sonora Elf Owl fit has slightly smaller $\log x_{tol}$ \change{than the ExoREM fit}, it requires slightly higher $b$, so we find that these fits are roughly equivalent in quality.

Both of the Sonora models prefer high temperature and small radius, while ExoREM prefers low temperature and large radius. We interpret this as being due to the presence of clouds in the ExoREM models, which limits the depth to which we can probe. This results in a lower temperature for cloudy models, which in turn results in a larger radius to conserve flux.

The fitted radius from the Sonora models is unphysically small for an object as massive as Ross 458c. The size of substellar objects on the order of a few Jupiter masses is limited by electron degeneracy pressure, with a minimum near $\sim1\ R_J$ \citep{Burrows2001}. For a comparatively young system such as Ross 458 ($< 1\rm\ Gyr$), one would expect the radius to be somewhat higher than $1\ R_J$. \change{The derived radius is about $0.75\ R_J$ based on the Sonora Bobcat models, $0.88\ R_J$ based on the Sonora Elf Owl models, and $1.3\ R_J$ based on the ExoREM models.} % If Ross 458c formed through core accretion in the disk, this restriction may not apply. For this reason, we cannot use the fitted radius to determine the preferred model if the formation pathway is still unknown.

We derive masses from the radius and $\log g$ values. From Sonora Bobcat, we derive a mass of $8.6\pm0.3\ M_J$. From Sonora Elf Owl, we derive a mass of $5.02\pm0.03\ M_J$, and from ExoREM, we derive a mass of $16.20\pm0.06\ M_J$. Each of these values is higher than the opacity limit for fragmentation, which makes them each consistent with a fragmentation scenario, and the ExoREM value is high enough to classify Ross 458c as a brown dwarf under the IAU working definition.

% Sonora Bobcat: $8.559^{+0.269}_{-0.270}\ M_J$
% Sonora Elf Owl: $5.015^{+0.032}_{-0.032}\ M_J$
% ExoREM: $16.196^{+0.064}_{-0.060}\ M_J$

Each of the grids prefers super-stellar metallicity models, \change{which would suggest formation} in a metal-enriched environment. The lowest metallicity model, ExoREM, is still higher than the metallicity we derived of the primary from the PHOENIX models, but it is neatly between the two estimates given by \cite{Burgasser2010}. The other two grids return a metallicity higher than any previous estimates for the host. The bulk of the evidence \change{from the forward model grids} suggests that the metallicity of Ross 458c is higher than that of its host.

Because we were unable to constrain the C/O ratio of the primary, we are limited in our ability to compare our fits of the companion. We can only say that our fitted C/O ratios for Ross 458c are \change{consistent with the observed range of C/O ratios in nearby M dwarfs, between 0.4 and 0.8 \citep{Nakajima2016}}.

% Using the relation from \citet{Nissen2013}, we estimate the C/O ratio of the primary to be $0.59\pm0.07$. The two model grids that fit for C/O, Sonora Elf Owl and ExoREM, return a best-fitting C/O within $1\upsigma$ of the primary. While Ross 458c appears to have enhanced metallicity relative to its host, the C/O ratio seems to be very similar.

\change{In summary, we find that while none of the forward model grids fully match the spectrum of Ross 458c, but the Sonora Elf Owl and ExoREM grids, which allow for vertical mixing and/or clouds, provide the best fits. Sonora Bobcat, which assumes a cloudless atmosphere in chemical equilibrium, requires higher error inflation than either Sonora Elf Owl or ExoREM to maximize its likelihood function, suggesting that the additional complexity of the latter grids is better able to characterize the spectrum. Still, Sonora Elf Owl and ExoREM require an additional uncertainty of close to 20\% of the flux, which we find to be unacceptable for model selection. If we adopt these models, we might interpret Ross 458c to have formed in a metal-rich environment, but first, we turn to atmospheric retrievals.} %Sonora Elf Owl, while requiring slightly similar error inflation to ExoREM, returns an unphysically small radius that is inconsistent with objects on the order of a few Jupiter masses. ExoREM is therefore our preferred model, suggesting that atmospheric complexity including clouds and vertical mixing are necessary to understand the structure of Ross 458c. The metallicities returned by these models are enhanced relative to the host, suggesting possible formation in a metal-enriched environment. %Sonora Elf Owl, while requiring slightly smaller error inflation than ExoREM, returns an unphysically small radius that is inconsistent with gaseous substellar objects. The ExoREM model is therefore our preferred model, and its metallicity and C/O ratio are consistent with the primary. Because its composition is consistent with the primary, the most likely formation scenario is gravitational fragmentation. We therefore conclude that Ross 458C formed as a brown dwarf companion to Ross 458AB.

% \begin{table*}
%     \begin{tabular}{|c|c|c|c|c|c|c|c|} \hline
%         Model Grid & Radius ($R_J$) & $T_{\rm eff}$ (K) & $\log g\rm\ (cgs)$ & [M/H] & C/O  & $\log K_{zz}\rm\ (cgs)$&$\log f$\\ \hline
%         Sonora Bobcat & $0.752^{+0.005}_{-0.005}$& $833.175^{+1.855}_{-1.720}$& $4.616^{+0.014}_{-0.013}$& $+0.450^{+0.011}_{-0.011}$& $0.458$ (fixed)  & $-$&$-0.422^{+0.004}_{-0.004}$\\ \hline
%  Sonora Elf Owl& $0.846^{+0.003}_{-0.003}$& $771.728^{+0.993}_{-0.988}$& $4.250^{+0.000}_{-0.000}$& $+0.572^{+0.010}_{-0.009}$& $0.596^{+0.004}_{-0.004}$& $3.151^{+0.083}_{-0.082}$&$-0.723^{+0.004}_{-0.003}$\\ \hline
%         ExoREM & $1.299^{+0.003}_{-0.004}$& $651.428^{+0.291}_{-0.333}$& $4.419^{+0.002}_{-0.002}$& $+0.258^{+0.007}_{-0.006}$& $0.639^{+0.001}_{-0.001}$& $-$&$-0.702^{+0.004}_{-0.003}$\\ \hline
%     \end{tabular}
%     \caption{Forward model results table}
%     \label{tab:fm_results}
% \end{table*}

\section{Atmospheric Retrievals}\label{sec:retrievals}

Atmospheric retrievals are an alternative method to fitting spectra of substellar companions \citep[e.g.][]{Madhusudhan2018}. Atmospheric retrieval codes typically generate millions of model spectra, spanning a range of atmospheric properties, and use Bayesian inference techniques like MCMC or Nested Sampling to explore the parameter space.

Retrieval models for directly-imaged substellar objects are typically defined by parameters encoding the pressure-temperature (P-T) profile, global properties (such as radius and gravity), chemical abundances, and aerosols \citep[e.g.][]{Line2015,Burningham2017,Piette2020,Nasedkin2024}. Atmospheric retrievals can  have much greater flexibility than forward model fitting because they have many more free parameters that allow them to explore departures from common forward model assumptions (e.g. radiative-convective and thermochemical equilibrium), but they are also potentially subject to unphysical results.

Due to the long computation time for atmospheric retrievals, we convolved and binned our spectrum to a resolving power of $\sim$300. To conserve the pixel spacing of the original spectrum, we convolved each of the spectra with a Gaussian whose FWHM is 9 times larger than the original FWHM, and we resampled the spectra with bin sizes increased by a factor of 9, so that the number of pixels per resolution element is conserved.

In this section, we provide atmospheric retrieval results using the open source retrieval code \texttt{POSEIDON}  \citep{Poseidon1, Poseidon2}. We outline the retrieval model setup in Section~\ref{sec:pconfig} and present the results in Section~\ref{sec:presults}.

\subsection{Retrieval Configuration} \label{sec:pconfig}

%\begin{table}
%    \caption{Line list references for the \Poseidon\ retrievals}
%    \centering
%    \begin{tabular}{lc} \hline \hline
%        Molecule& Source \\ \hline
%        H$_2$O& \cite{H2O} \\
%        CH$_4$& \cite{CH4} \\
%        CO& \cite{CO} \\
%        CO$_2$& \cite{CO2} \\
%        H$_2$S& \cite{H2S} \\
%        NH$_3$& \cite{NH3} \\
%        K& \cite{NaK} \\ \hline
%    \end{tabular}
%    \label{tab:molecules}
%\end{table}

\begin{table}
    \caption{Free parameters and priors used for the \Poseidon\ retrievals}
    \centering
    \begin{tabular}{llc} \hline\hline
         Parameter & Description & Priors \\ \hline
         ${\rm R_{p, ref}}$ & \change{Planet reference radius at 10\,bar (R$_{\rm Jup}$)} & $\mathcal{U}(0.46, 4.6)$ \\
         $\log_{10} g$ & \change{Gravity at reference radius (cgs)} & $\mathcal{N}(4.5, 0.7^2)$\\
         $d$ & \change{System distance (pc)} & $\mathcal{N}(11.51, 0.02^2)$ \\
         $\log_{10} X_i$ & \change{Volume mixing ratio of i$^{\rm{th}}$ gas} & $\mathcal{U}(-12,-1)$\\
         $T_{\rm phot}$ & \change{Temperature at $10^{0.5}$\,bar (K)} & $\mathcal{U}(100, 3000)$\\
         $\Delta T_{1}$ & \change{Temperature difference from $10^{-3}$--$10^{-2}$\,bar (K)} & $\mathcal{U}(0, 1000)$ \\
         $\Delta T_{2}$ & \change{Temperature difference from $10^{-2}$--$10^{-1}$\,bar (K)} & $\mathcal{U}(0, 1000)$ \\
         $\Delta T_{3}$ & \change{Temperature difference from $10^{-1}$--$10^{0}$\,bar (K)} & $\mathcal{U}(0, 1000)$ \\
         $\Delta T_{4}$ & \change{Temperature difference from $10^{0}$--$10^{0.5}$\,bar (K)} & $\mathcal{U}(0, 1000)$ \\
         $\Delta T_{5}$ & \change{Temperature difference from $10^{0.5}$--$10^{1}$\,bar (K)} & $\mathcal{U}(0, 1500)$ \\
         $\Delta T_{6}$ & \change{Temperature difference from $10^{1}$--$10^{1.5}$\,bar (K)} & $\mathcal{U}(0, 2000)$ \\
         $\Delta T_{7}$ & \change{Temperature difference from $10^{1.5}$--$10^{2}$\,bar (K)} & $\mathcal{U}(0, 2500)$ \\
         $\log_{10} P_{\rm{cloud}}$ & \change{Cloud top pressure (bar)} & $\mathcal{U}(-3, 2)$ \\
         $\Delta \log_{10} P_{\rm{cloud}}$ & \change{Cloud thickness (bar)} & $\mathcal{U}(0, 5)$ \\
         $\log_{10} r_m$ & \change{Aerosol mean particle radius ($\upmu$m)} & $\mathcal{U}(-3, 1)$ \\
         $\log_{10} X_{\rm{aerosol}}$ & \change{Aerosol volume mixing ratio} & $\mathcal{U}(-20, -1)$ \\
         $b$ & \change{Error inflation exponent} \citep{Line2015} & $\mathcal{U}(\log_{10} (10^{-3} \min[\sigma^2]), \log_{10} (10^{2} \max[\sigma^2]))$ \\
         $x_{tol}$ & \change{Fractional error inflation} \citep{Piette2020} & $\mathcal{U}(0.051, 1)$  \\ \hline
    \end{tabular}
    
    \label{tab:retrieval_priors}
\end{table}

\change{Our retrieval analysis for Ross 458c sees one of the first applications of \Poseidon\ to the atmospheric analysis of a directly imaged substellar object emission spectrum. While \Poseidon\ has been widely used to interpret exoplanet transmission spectra \citep[e.g.][]{Poseidon1, MacDonald2019, Louie2024, B-A2025} and, more recently, to exoplanet secondary eclipse spectra \citep[e.g.][]{Coulombe2023, Gressier2025}, \Poseidon\ v1.2 adds support for emission spectra retrievals of directly imaged object emission spectra with multiple scattering \citep{Mullens2024}. The underlying forward model with thermal scattering is adapted from the open source code \texttt{PICASO} \citep{Batalha2019, Mukherjee2023}, with Mie scattering aerosols included as described in  \citet{Mullens2024}.}

Our retrieval model assumes a H$_2$-He background gas (using the solar ratio of $\rm He/H_2 = 0.17$ \citealt{Asplund2009}), and includes line opacity from the following chemical species: H$_2$O \citep{H2O}, CH$_4$ \citep{CH4}, CO \citep{CO}, CO$_2$ \citep{CO2}, H$_2$S \citep{H2S}, NH$_3$ \citep{NH3}, and K \citep{NaK}. \change{Continuum opacity from H$_2$-H$_2$ and H$_2$-He collision induced absorption is included from HITRAN \citep{Karman2019}, alongside and H$_2$ Rayleigh scattering \citep{Hohm1994}. We consider four potential aerosol species in this work (one aerosol species in each retrieval model), with the aersols modeled as homogeneous Mie scattering spheres \citep{Mullens2024}: Na$_2$S \citep{Morley2012,Wakeford2015}, MgSiO$_3$ (crystalline) \citep{Jaeger1998,Burningham2021}, KCl \citep{Palik1985,Wakeford2015}, and MnS \citep{Huffman1967,Wakeford2015}.} The pressure grid of our atmosphere models range from $10^{-3}$ to $10^2$ bar, with 100 layers uniformly spaced in log-pressure. We calculate emission spectrum models \change{using opacity sampling from high-resolution ($\Delta\nu = $ 0.01\,cm$^{-1}$, or $R = \lambda/\Delta\lambda = 10^6$ at 1\,$\upmu$m) cross sections} onto a wavelength grid from 0.8--3.2\,$\upmu$m at a spectral resolution of $R = \lambda/\Delta\lambda = 30,000$, before binning these models down to $R\sim300$ match the data. We use a spectral resolution a factor of 100 higher than the data to minimize errors from the sampling resolution.

\change{Our \Poseidon\ retrieval model spans 24 free parameters, which encode a range of physical, chemical, thermal, and aerosol properties. The physical and system properties are the planetary radius at the 10\,bar reference pressure, the log$_{10}$-surface gravity at the reference radius, and the distance to the system. We place Gaussian priors on the system distance from Gaia ($\mathcal{N}(11.51, 0.02^2)$\,pc; \citealt{Stassun2019}) and the log$_{10}$-surface gravity at ($\mathcal{N}(4.5, 0.7^2)$) in cgs units, consistent with the $\log g$ range from previous work (see Table \ref{tab:literature}) and wide enough to give the retrieval model flexibility. We prescribe an uninformative uniform prior on the planetary radius ($\mathcal{U}(5, 50) R_E = \mathcal{U}(0.46, 4.6) R_J$). The log$_{10}$-volume mixing ratios of the 7 gases described above --- assumed constant with altitude --- span an uninformative wide prior range ($\mathcal{U}(-12, -1)$). The P-T profile follows the `slope' prescription from \citet{Piette2020}, which parametrizes a monotonically increasing temperature with increasing pressure via a reference photosphere temperature and 7 temperature differences over different pressure intervals. We parametrize aerosols using the `slab' model from \citet{Mullens2024}, which is defined by the cloud top pressure, the cloud thickness, the mean aerosol particle radius (assuming a log-normal particle size distribution), and the aerosol log$_{10}$-volume mixing ratio. Finally, we consider the same two sources of error inflation as Section~\ref{sec:host_z}: wavelength-independent inflation (accounting for uncertainty underestimation in the data) and proportional-to-flux inflation (accounting for model uncertainties --- see below). A full list of the free parameters and adopted priors are shown in Table~\ref{tab:retrieval_priors}. All retrievals sample the parameter space using 400 MultiNest \citep{Feroz2009, Buchner2014} live points.
}

\begin{figure*}
    \centering
    \includegraphics[width=0.49\linewidth]{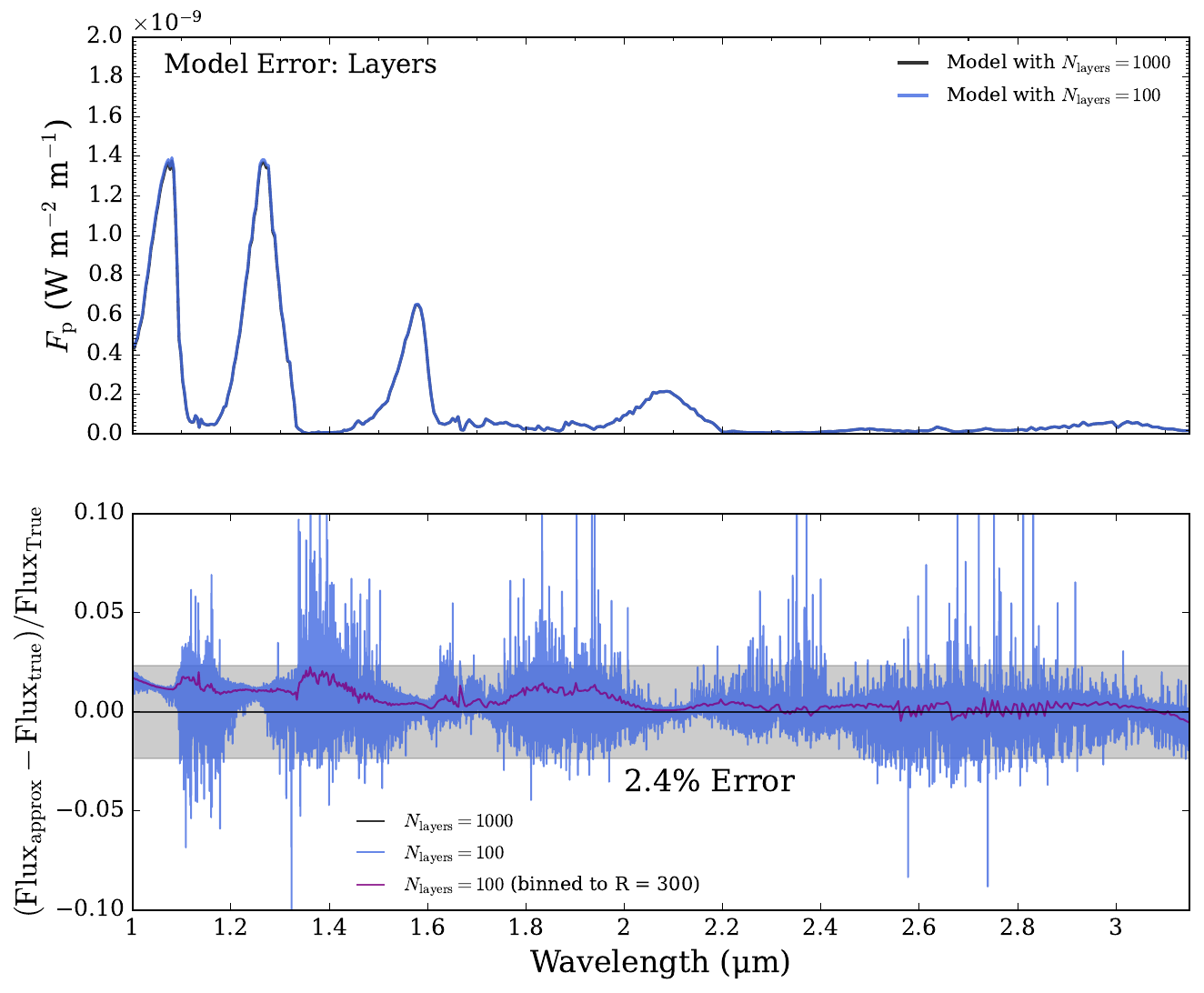}
    \includegraphics[width=0.49\linewidth]{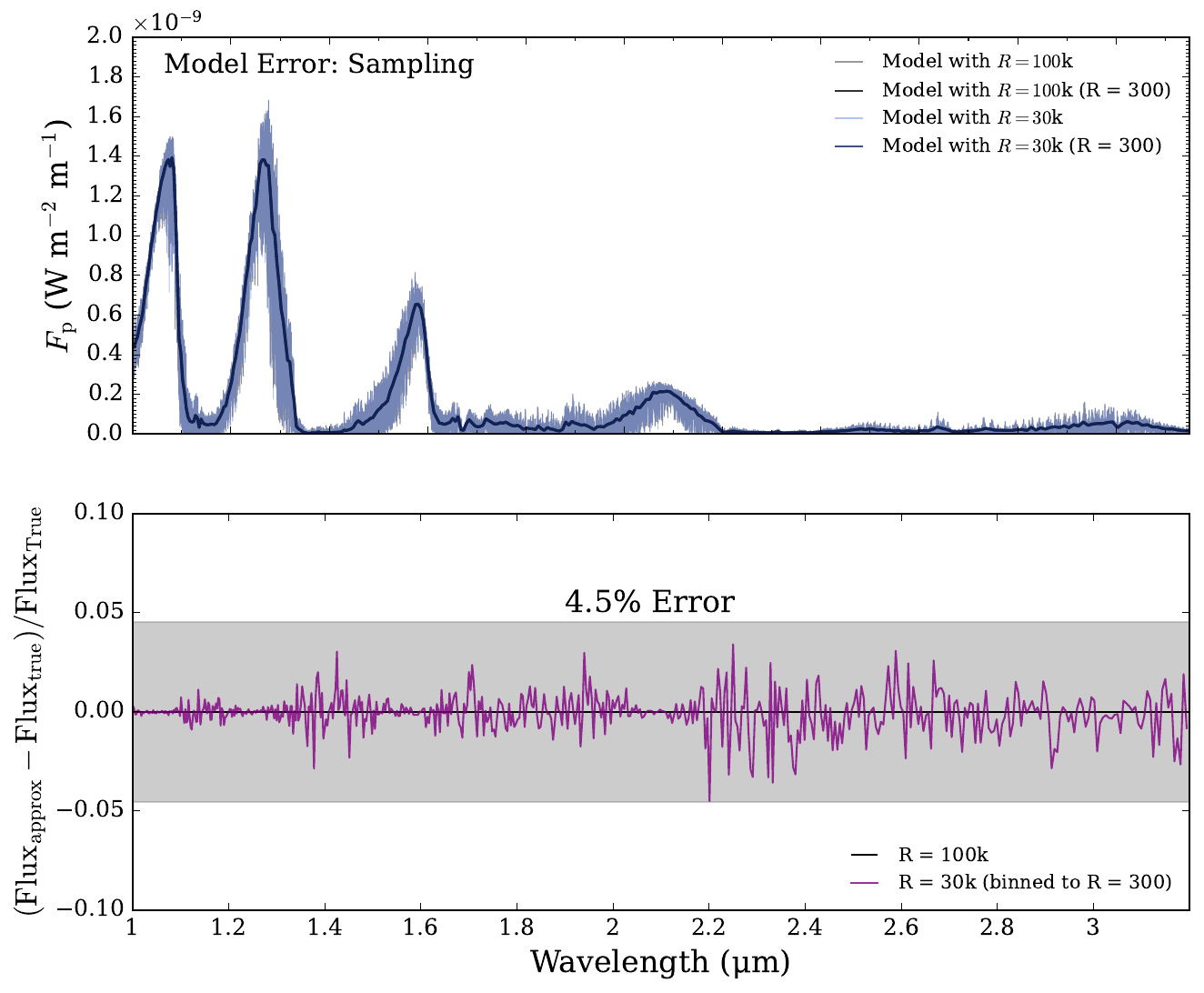}
    \caption{\change{Sources of model error for the \Poseidon\ retrievals. Left: model spectra with 100 layers vs. 1,000 layers (top panel) impart a maximum wavelength-dependent relative flux error of 2.4\% after binning to $R = 300$ (bottom panel). Right: model spectra with opacity sampling at $R_{\rm{model}} =$ 30,000 vs. 100,000 (top panel) impart a maximum wavelength-dependent relative flux error of 4.5\% after binning to $R = 300$ (bottom panel). These two errors combine in quadrature to a model error of 5.1\%.}}
    \label{fig:model_errors}
\end{figure*}

\change{We account for model uncertainty in our atmospheric retrievals by considering two sources of error: (i) the finite number of atmospheric layers; and (ii) opacity sampling onto the adopted model wavelength grid. We use these two error estimates to place an appropriate lower-limit on the proportional-to-flux `$x_{tol}$' parameter, as described in \citet{Piette2020}. First, we show in Figure~\ref{fig:model_errors} (left panels) that the maximum relative flux error associated with using 100 atmospheric layers (as adopted in our retrieval models), relative to a reference model with 1,000 atmospheric layers, is 2.4\% after binning the model to $R = 300$ (the resolution of the data in our retrievals). A similar exercise for opacity sampling is shown in Figure~\ref{fig:model_errors} (right panels), for which the sampling error from using an $R_{\rm{model}} =$ 30,000 grid yields a maximal error of 4.5\% relative to a reference model at $R_{\rm{model}} =$ 100,000. Combining these two error sources in quadrature yields an estimated model error of 5.1\%, which we set as the minimum prior for the $x_{tol}$ parameter. This lower prior allows the posterior distributions to broaden to reflect these known model errors from radiative transfer that will always be present in our retrievals.}

\change{After running retrievals, we calculate posteriors for the elemental ratios from the retrieved volume mixing ratios. For example, the atmospheric O/H and C/O ratios are given by:}
\begin{equation}
    \mathrm{O/H} = \frac{X_{\mathrm{H_2 O}} + X_{\mathrm{CO}} + 2X_{\mathrm{CO_2}}}{2X_{\mathrm{H_2}} + 2X_{\mathrm{H_2 O}} + 4X_{\mathrm{CH_4}} + 2X_{\mathrm{H_2 S}} + 3X_{\mathrm{NH_3}}}
\end{equation}
\begin{equation}
    \mathrm{C/O} = \frac{X_{\mathrm{CO}} + X_{\mathrm{CO_2}} + X_{\mathrm{CH_4}}}{X_{\mathrm{H_2 O}} + X_{\mathrm{CO}} + 2X_{\mathrm{CO_2}}}
\end{equation}
\change{We express our results for O/H, C/H, N/H, and S/H relative to the solar abundances from \citet{Asplund2021} on a logarithmic scale, such that [O/H] = $\log_{10} \left(\rm{\frac{(O/H)_{atm}}{(O/H)_{solar}}} \right)$. Finally, the atmospheric metallicity is given by:}
\begin{equation}
    \rm{[M/H] = \log_{10} \left( \frac{(O/H)_{atm} + (C/H)_{atm} + (N/H)_{atm} + (S/H)_{atm}}{(O/H)_{solar} + (C/H)_{solar} + (N/H)_{solar} + (S/H)_{solar}} \right)}
\end{equation}

\begin{deluxetable}{lcccccccc}
    \tablecaption{Sensitivity of retrieved elemental ratios to different cloud models}
    \tablewidth{0pt}
    \tablehead{
    Cloud Model & \colhead{$\ln \rm \mathcal{Z}$} & \colhead{[O/H]} & \colhead{[C/H]} & \colhead{[N/H]} & \colhead{[S/H]} & \colhead{[M/H]} & \colhead{C/O}
    }
    \startdata
        Na$_2$S & 24223 & $+0.05^{+0.03}_{-0.02}$ & $+0.11^{+0.04}_{-0.03}$ & $-1.23^{+0.06}_{-0.06}$ & $+0.41^{+0.08}_{-0.09}$ & $+0.05^{+0.04}_{-0.02}$ & $0.68^{+0.03}_{-0.02}$ \\
        MgSiO$_3$ (Crystalline) & 24196 & $+0.06^{+0.03}_{-0.02}$ & $+0.13^{+0.04}_{-0.03}$ & $-1.17^{+0.06}_{-0.06}$ & $+0.37^{+0.09}_{-0.10}$ & $+0.06^{+0.03}_{-0.02}$ & $0.69^{+0.03}_{-0.02}$ \\
        KCl & 24179 & $+0.08^{+0.11}_{-0.04}$ & $+0.13^{+0.16}_{-0.05}$ & $-1.17^{+0.06}_{-0.07}$ & $+0.39^{+0.09}_{-0.12}$ & $+0.07^{+0.12}_{-0.04}$ & $0.68^{+0.07}_{-0.04}$ \\
        MnS & 24102 & $+0.07^{+0.07}_{-0.03}$ & $+0.11^{+0.11}_{-0.04}$ & $-1.15^{+0.05}_{-0.08}$ & $+0.69^{+0.05}_{-0.06}$ & $+0.08^{+0.08}_{-0.04}$ & $0.65^{+0.06}_{-0.02}$ \\
    \enddata
    \tablecomments{Models are sorted by the Bayesian evidence ($\mathcal{Z}$), with the best fitting model at the top. The atmospheric elemental abundances are expressed relative to solar abundances, e.g., $\rm{[O/H] = \log_{10} \left( (O/H)_{atm} / (O/H)_{solar} \right)}$. The metallicity, [M/H], accounts for all the elements heavier than He relative to their solar abundances. C/O is expressed in absolute units rather than a logarithmic scale.}
    \label{tab:poseidon_results}
\end{deluxetable}

%\texttt{POSEIDON} is an atmospheric retrieval code for modeling transit and emission spectra \citep{Poseidon1, Poseidon2}.

\subsection{Retrieval Results} \label{sec:presults}

\change{Table~\ref{tab:poseidon_results} summarizes the derived elemental ratios for each of our cloud models, ranked by their Bayesian evidence (calculated by MultiNest). Since the model with Na$_2$S clouds has the highest Bayesian evidence, we focus primarily on this model moving forward. However, we note that the retrieved atmospheric composition is consistent (within 1\,$\sigma$), regardless of the adopted cloud model.}

Our retrieved atmospheric properties for Ross 458c are summarized in Figure \ref{fig:poseidon}, \change{and the full parameter distribution is presented in Appendix \ref{app:corner}}. Looking at the spectrum, we see a much better fit than the forward models. We find that the model is able to reproduce the shape of each major peak in the spectrum, and the only major discrepancy is at shorter wavelengths than the Y-band. Alkali opacities dominate in this regime, and these are not fully understood \citep{Phillips2020}.

The pressure-temperature profile \change{varies smoothly}, and we are able to constrain H$_2$O, CO$_2$, CH$_4$, H$_2$S, NH$_3$, and K. CO is not detected in the spectrum.

\change{Our best-fitting model returns an atmospheric metallicity of $\rm [M/H] = +0.05^{+0.04}_{-0.02}$, within $1\upsigma$ of the primary. Each of the other models also returns a metallicity within $1\upsigma$ of the primary, indicating that this result is robust against choice of cloud species.} The C/O ratio is slightly supersolar, but within the range observed for nearby M dwarfs, \change{between 0.4 and 0.8} \citep{Nakajima2016}. Because we did not fit for the C/O ratio of the primary, \change{we cannot state conclusively if it is consistent with the primary.}

\change{We find that the retrieved abundances of nitrogen and sulfur deviate somewhat from the solar abundance, but as these elements are much less abundant than oxygen and carbon, they have little effect on the overall metallicity. The metallicity is primarily driven by the oxygen and carbon abundances.}

\change{We also note that we only report the atmospheric metallicity from the retrieved abundances of molecules, and does not account for any metals in clouds. Because our best-fitting model is the Na$_2$S cloud model, and sodium and sulfur are trace relative to carbon and oxygen, we do not expect this to impact the conclusions regarding the overall atmospheric metallicity.}

\begin{figure*}
    \centering
    \includegraphics[width=1\linewidth]{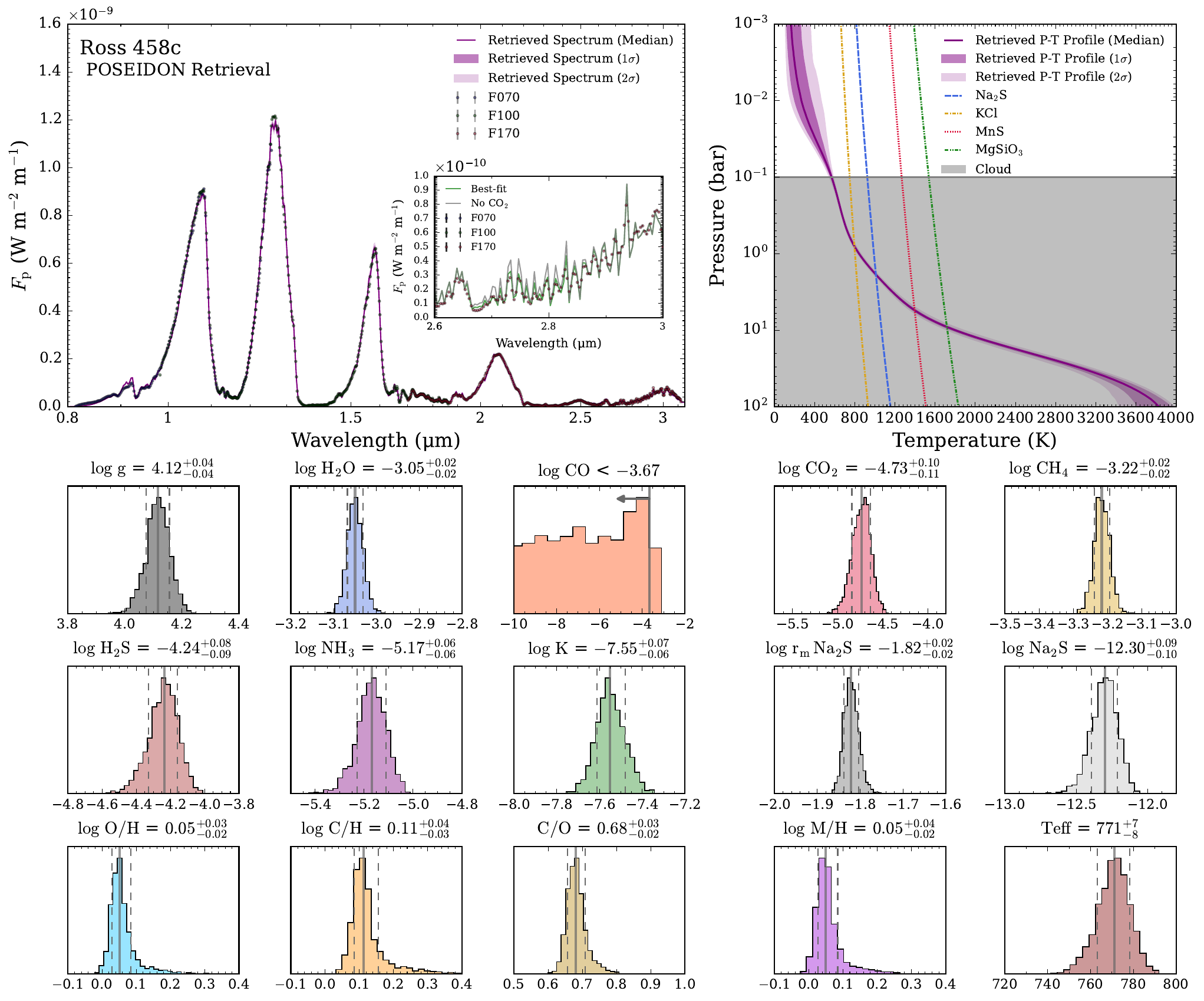}
    \caption{Summary figure of the fit to Ross 458c using \Poseidon\ atmospheric retrievals. Top left: median, $1\upsigma$, and $2\upsigma$ retrieved spectra compared to the data. Inset is the wavelength range of a 2.8 \textmu m CO$_2$ absorption band, showing a model with and without CO$_2$. Top right: retrieved pressure-temperature profile, with condensation curves of the cloud species considered overplotted, from \texttt{Virga} \citep{Virga}, and the retrieved cloud location shaded gray. Bottom rows: probability distributions of key parameters, including $\log g$, molecular abundances, cloud properties, C/O, and [M/H].}
    \label{fig:poseidon}
\end{figure*}

% \subsection{\texttt{BREWSTER}}

% \subsubsection{Retrieval Configuration}
% \note{Describe \texttt{BREWSTER} here.}

% \subsubsection{Results}
% \note{Describe \texttt{BREWSTER} results here.}

% \subsection{\texttt{APOLLO}}

% \subsubsection{Retrieval Configuration}
% \textcolor{magenta}{Describe \texttt{APOLLO} here.}

% \subsubsection{Results}
% \textcolor{magenta}{Describe \texttt{APOLLO} results here.}

\begin{table*}
    \centering
    \begin{tabular}{cccccc}\hline\hline
         Parameter&  Units&  Sonora Bobcat&  Sonora Elf Owl&  ExoREM& \Poseidon\\ 
          & & Clear & Clear & Clouds & Na$_2$S Clouds \\
          & & Eq. Chem. & Diseq. Chem. & Diseq. Chem. & Free Chemistry \\\hline
         Radius&  $R_J$&  0.75&  0.88&  1.3&$0.85^{+0.02}_{-0.02}$\\
 Mass& $M_J$& 8.6& 5.0& 16&$8.24^{+0.80}_{-0.75}$\\
         $T_{\rm eff}$&  K&  830&  760&  650&$771.1^{+7.2}_{-8.0}$\\
         $\log g$&  cgs&  4.6&  4.3&  4.4&$4.12^{+0.04}_{-0.04}$\\
         $\rm [M/H]$&  $-$&  +0.45&  +0.50&  +0.25&$+0.05^{+0.03}_{-0.02}$\\
         C/O&  $-$&  $-$&  0.61&  0.64&$0.68^{+0.03}_{-0.02}$\\ 
         $\chi_{\rm red}^2$ (no err. infl.) & $-$& 576& 212& 159& $-$\\
         $\chi_{\rm red}^2$ (w/ err. infl.) & $-$& 1.36& 1.19& 1.47& 1.00\\ \hline
    \end{tabular}
    \caption{Results of Ross 458c parameters from the forward model grid comparisons and the \Poseidon\ retrieval with Na$_2$S clouds. For the forward models, because the uncertainties are too tightly constrained, we report parameters to two significant figures. We do not report the $\chi_{\rm red}^2$ without error inflation for the \Poseidon\ retrieval because we impose a minimum $x_{tol}$ of 5.1\% determined by the uncertainty in the model, which inherently limits the minimization of $\chi_{\rm red}^2$.}
    \label{tab:results}
\end{table*}

\section{Discussion} \label{sec:discussion}

\subsection{Comparison to other work}

All of the forward models prefer an enhanced metallicity for Ross 458c relative to its host. There is growing evidence that the same is true of 51 Eri b and HR 8799b, suggesting there is a new class of widely separated planets with enriched metallicity \citep{Balmer2025, Nasedkin2024}.

The \Poseidon\ retrievals, on the other hand, uniformly prefer a metallicity consistent with the primary within $1\upsigma$. This would suggest that Ross 458c is similar to the majority of brown dwarf companions, which have the same composition as their host \citep{Hoch2023, Xuan2024}. \change{The metallicity is also consistent with previous work by \cite{Goldman2010} within $1\upsigma$ and \cite{Zalesky2022} within $2\upsigma$. The [C/H] ratio is consistent with the retrieval by \cite{Gaarn2023}, but they find Ross 458c to be oxygen-depleted, which we do not find here.}

Although we do not directly compare the C/O ratio of Ross 458c with the host, we retrieve a value of $0.68^{+0.03}_{-0.02}$, which is consistent with nearby M dwarfs \citep{Nakajima2016}. We find disagreement with the findings of \citet{Gaarn2023}, who found that the C/O ratio of Ross 458c is close to 2. We hypothesize that the result from \citet{Gaarn2023} may have resulted from older line lists for carbon- and oxygen-bearing species, which could introduce differences in the retrieved abundances that would propagate into the derived C/O ratio. State-of-the-art molecular line lists are constantly being updated, and the ExoMol line list for CH$_4$ was updated as recently as 2024 \citep{CH4}. Further investigation will be needed to determine if this is the cause of their \change{unusually high} C/O ratio.

%\citet{Balmer2025} finds that 51 Eri b, a 0.88 $M_J$ companion at 9.58 AU from its host is best fit by a model with [M/H] = +0.65 and C/O = 0.65, a modest departure from solar in C/O, but a significant departure from its [Fe/H] = -0.1 host \citep{Koleva2012, Rajan2017}. Similarly, \citet{Nasedkin2024} finds that HR 8799b, a $6\ M_J$ companion at 71 AU from its host, has a metallicity [M/H] = $+0.96^{+0.08}_{-0.08}$ and a C/O ratio of $0.78^{+0.03}_{-0.04}$. This is a slightly greater departure in C/O than 51 Eri b, but a much greater departure from its subsolar metallicity host \citep{Moya2010}. Ross 458c adds to a growing sample of metal-enriched widely separated planetary mass companions.

%\note{Compare to GJ 504b? b is metal rich, but I'm pretty sure A is too. Latest analysis of A was line-by-line which seems less goofy than the forward model interpolator I'm doing.}

%We find that our C/O values agree with the findings of \citet{Hoch2023}, which found that widely separated companions tend to concentrate around the solar value of C/O. \citet{Xuan2024} compared [M/H] and C/O for a sample of widely separated companions, and found that they also concentrate around the solar values. Ross 458c would be on the higher end of their sample, but not outside their range of values. \note{Don't really know how to describe this without just using their figure and putting a dot where Ross 458c goes. Can I do that?}

\subsection{Implications for formation scenario}

Using the forward models, we find metal-enrichment for Ross 458c relative to its host. The range of metal enrichment ranges from 1.6 to 2.9 times solar, depending on the model, similar to the metal enrichment of Jupiter. This could either be due to metal accretion or hydrogen loss. Some combination of accreting carbon-rich material and icy material could increase the metallicity of Ross 458c \citep{KemptonKnutson2024}. Alternatively, Ross 458c may have lost material that was preferentially hydrogen and helium. \citet{Nayakshin2014} suggests a core-assisted gas capture model which would lead to a core that is preferentially metal-rich, and an envelope that is preferentially metal-poor. In this scenario, if Ross 458c were to lose any mass to the disk, it would be from the metal-poor envelope, leading to an enhanced metallicity. Either of these scenarios would require Ross 458c to have been in a protoplanetary disk in order to exchange materials with its surroundings, before migrating outward to its current separation.

The retrievals, on the other hand, indicate that the metallicity of the companion is consistent with the metallcity of the host. This suggests that Ross 458c likely formed through \change{a ``top-down" formation process}. Because the atmospheric retrieval returns a much better fit than the forward models, we \change{adopt the retrieval model and} conclude that Ross 458c is most likely a brown dwarf companion to Ross 458AB that formed through fragmentation.

\subsection{Detection of CO$_2$}

We are able to constrain CO$_2$ in the atmosphere of Ross 458c, at an abundance of $\log \rm CO_2 = -4.73^{+0.10}_{-0.11}$. This is an unexpected result for a solar metallicity object in the brown dwarf temperature regime. CH$_4$ is expected to dominate the abundance ratios at solar metallicity, while CO$_2$ becomes more important at high metallicity \citep{Moses2013B}. In addition, CO$_2$ does not have many strong absorption features in the near IR. There is a weak feature at 2.8 \textmu m, but the dominant CO$_2$ absorption band from 1$-$5 \textmu m is at about 4.3 \textmu m \citep{CO2}.

The 2.8 \textmu m feature has been detected in transmission spectra with JWST \citep{Feinstein2023, Taylor2023, FT2024}, and it appears as if this is where we detect CO$_2$ with the \Poseidon\ retrieval as well. The inset in the top-left panel of Figure \ref{fig:poseidon} shows the best-fitting model with and without CO$_2$ from 2.6$-$3 \textmu m, and we find that CO$_2$ provides noticeable absorption in this wavelength range.

Other studies have found an overabundance of CO$_2$ compared to forward models in objects in a similar temperature range as considered here \citep[e.g.][]{Lew2024, Beiler2024}. Observations of Ross 458c at longer wavelengths will be needed to confirm if this CO$_2$ detection is real or not. A NIRSpec IFU G395H spectrum of Ross 458c has been taken as part of JWST GTO program \#1277 (PI: P. Lagage), and we look forward to seeing if they are able to confirm this CO$_2$ detection.

\subsection{Forward models vs. atmospheric retrievals}

Forward models of substellar objects are useful for getting an estimate of their effective temperatures and radii because the small number of parameters allows for the parameter space to be explored relatively quickly. We find, however, that it is difficult to constrain some of the more subtle parameters, like metallicity and C/O, because the fits are poor. Additionally, there is little to no advantage to using an interpolated grid of forward models over a simple chi-squared minimization. Because of the high signal-to-noise, the parameter distributions are too small to be physically reasonable. We attempted to account for this by using error inflation terms to capture the error in the models, but our distributions are still too tight and vary too much from model to model to be considered \change{reasonable}.

Atmospheric retrievals are our preferred method of fitting high-resolution, high signal-to-noise emission spectra. Retrievals allow for variation of specific molecules, which removes the restrictions imposed by forward models through their parameterization of bulk metallicity and C/O. Substellar companions are complex objects, and their complexity can only be captured by a method which allows for varying abundances of individual molecules.

Retrievals are still, of course, not perfect. We found that there is an inherent error corresponding to 5.1\% of the flux in the forward models generated by \Poseidon\ due to the finite opacity sampling and number of pressure layers in the radiative transfer code. We account for this using error inflation terms that capture the uncertainty in the forward models and the opacities, but care must be taken to ensure that retrieval results are physically reasonable.

\subsection{The metallicity bias of forward models}

A surprising result from our forward model analysis is that all of our forward models preferred high metallicity (by a factor between +0.20 and +0.45 dex, depending on the model). They were all poor fits, so it is ill-advised to draw any conclusions based on the results. However, if they were unable to accurately reproduce the metallicity, there is no obvious reason why they would each tend towards high metallicity. Due to the many degeneracies between metallicity, pressure-temperature profile, clouds, and other assumptions, more work is needed to investigate this phenomenon. We find that this method returns a much better fit for the primary than for the companion, so we do not expect the metallicity of the primary to be biased in the same way.

\section{Conclusions} \label{sec:conclusions}

By comparing the PHOENIX grid of stellar models with a near-IR spectrum of Ross 458AB, we found that Ross 458AB is less metal-rich than previously assumed, and its metallicity is much closer to solar. Constraining the metallicity of the host is essential to any study of the formation of a companion.

Using three forward model grids, we found that Ross 458c has enhanced metallicity relative to its host, but the fits were all too poor to draw robust conclusions. Using the \Poseidon\ retrieval code, we found that its metallicity is actually consistent with its host within 1$\upsigma$, and returned a much more reasonable fit compared to the forward models. Based on these results, we conclude that Ross 458c has a composition very similar to its host, indicating formation through fragmentation.

We find that forward models are insufficient to capture the complexity of high-resolution emission spectra of low-mass T dwarfs. With interpolated grids of forward models, one cannot control the exact abundances of each molecule, and so the parameter space is too limited for accurate characterization. We need atmospheric retrievals to adequately characterize the atmospheres of directly imaged substellar objects.

Future steps will be to compare our \Poseidon\ retrievals with the publicly available retrieval codes \Brewster\ \citep{Brewster} and \Apollo\ \citep{Apollo}, to see if our results are robust against choice of retrieval code. \change{In addition, we plan to try more complex models that allow for inhomogeneous atmospheres, with patchy clouds, multiple cloud species, and a vertical methane chemical gradient. \citet{Zhang2025-TWA27B} found that choices in modeling a homogeneous or inhomogenous atmosphere can lead to vast differences in retrieved metallicities and C/O. \citet{Manjavacas2019} found that Ross 458c has a variable white light curve at about the 2.6\% level, suggesting an inhomogeneous atmosphere. The period of this variability is about 6 hours, which is close to the total time of our observation. This suggests that the three spectra used in this work were taken during different points in the object's light curve. By calibrating the spectra with each other and adjusting the errors based on the statistical uncertainty in the individual frames, we account for some of the variability, but more nuanced retrievals will be required to fully disentangle its inhomogeneity.}

\change{Modeling these additional complexities will help us better characterize Ross 458c. So far, we have found that the composition of Ross 458c is consistent with its host for every choice of cloud species when we assume a homogeneous atmosphere. If this consistency holds for every choice of retrieval code and atmospheric model, it would be further evidence that the composition of Ross 458c matches its host, suggesting formation through fragmentation.}

% There are many additional steps we can take to further characterize the atmosphere of Ross 458c. We plan to try retrievals with patchy clouds, in which the spectrum is a linear combination of a cloud model and a clear model, parameterized by a filling fraction. Additionally, we plan to try using multiple cloud species and a vertical chemical gradient for methane, one of the dominant molecules in the atmosphere of Ross 458c. We will see if these additional complexities improve the fit, or if the new parameters have little to no effect on the spectrum. Finally, more data will be essential to understanding the complex atmospheres of late T dwarfs like Ross 458c.

\section*{Acknowledgments}
This work is based in part on archival data obtained with the NASA Infrared Telescope Facility, which is operated by the University of Hawaii under a contract with the National Aeronautics and Space Administration. Part of this work was carried out at the Jet Propulsion Laboratory, California Institute of Technology, under a contract with the National Aeronautics and Space Administration. \copyright\ 2025. All rights reserved. \change{We thank the anonymous referee for their helpful feedback that improved the quality of this study.}

\appendix

\section{Ross 458AB joint fit} \label{app:joint}

\begin{figure*}
    \centering
    \includegraphics[width=1\linewidth]{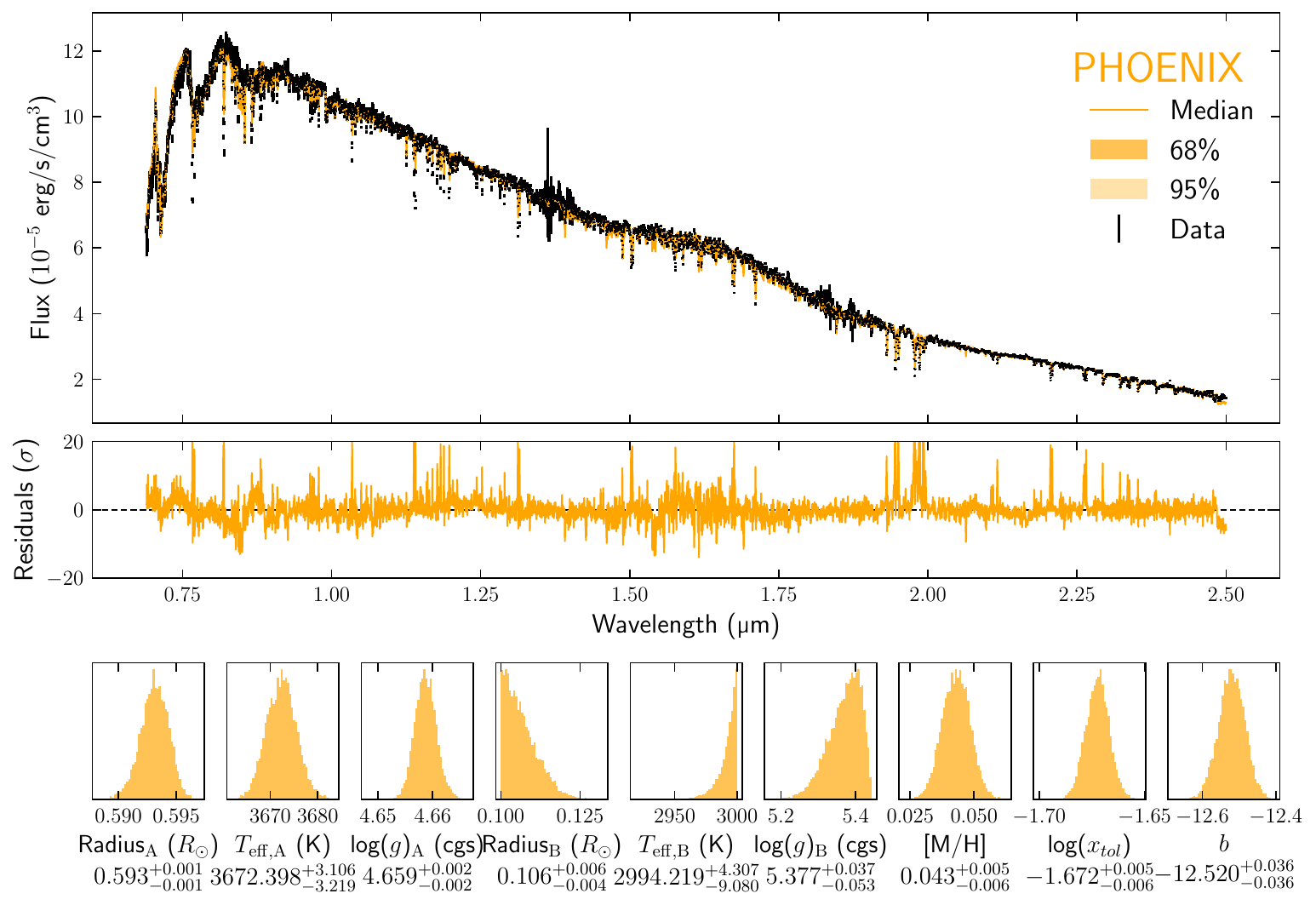}
    \caption{Summary figure of the fit to Ross 458AB using interpolated grids of PHOENIX models. The description of each panel is the same as in Figure \ref{fig:phoenix}, but here we fit to both the A and B components. We use the same metallicity for both components.}
    \label{fig:joint}
\end{figure*}

We attempt to fit Ross 458AB using both components of the M0.5+M7 binary, and show that the flux from the secondary is too small to be accounted for in the spectrum of the binary. We follow the same process as in Section \ref{sec:host_z}, but allowing for a separate A component and B component for radius, temperature, and $\log g$. We use the same priors for A as we used in Section \ref{sec:host_z}. For B, we use priors determined by comparing to evolutionary tracks for a 0.09 $M_\odot$ star from AMES-Cond \citep{Allard2001, Baraffe2003}. We use the same metallicity for both components because we assume that a tight binary should have the same composition. For each step in the \texttt{emcee} run, we query a spectrum for A and B from our interpolated PHOENIX grid and add them together to form a model spectrum, which we then compare to the host. We show a summary plot of this run in Figure \ref{fig:joint}.

\begin{figure}
    \centering
    \includegraphics[width=0.75\linewidth]{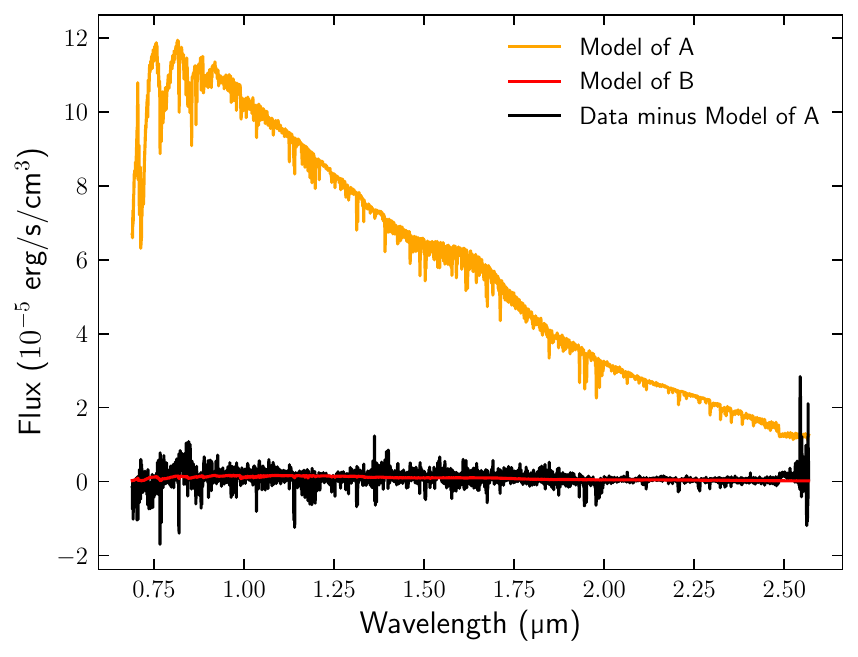}
    \caption{Model of Ross 458A and Ross 458B from the median returned parameters from the joint fit. We also plot the SpeX data subtracted by the model of Ross 458A to show that the remaining flux fitted by Ross 458B is just scatter around 0.}
    \label{fig:AandB}
\end{figure}

Looking at the returned values for $\log x_{tol}$ and $b$, we find virtually no improvement in the fit, indicating that the joint fit still required the same degree of error inflation as the fit that just used the A component. We find that the radius of B is hitting the edge of the prior, indicating that the model prefers to squash the flux from B entirely. In Figure \ref{fig:AandB}, we query the median model of A and B from our interpolators, and compare the model of B to the difference of the data and the model of A.

The leftover contribution from B is negligible compared to the contribution from A. When we fit both spectra, the model prefers to use the flux from A to fit the spectrum, and what remains is just scatter around 0. The returned radius of B is pushing the lower edge of the radius prior because the model is attempting to fit it to scatter around 0. Additionally, whether we include B or not, we find that the metallicity of the host is consistent within 1$\upsigma$. For these reasons, we chose to only model Ross 458A in Section \ref{sec:host_z}.

\section{\Poseidon\ corner plot}\label{app:corner}

\begin{figure}
    \centering
    \includegraphics[width=\linewidth]{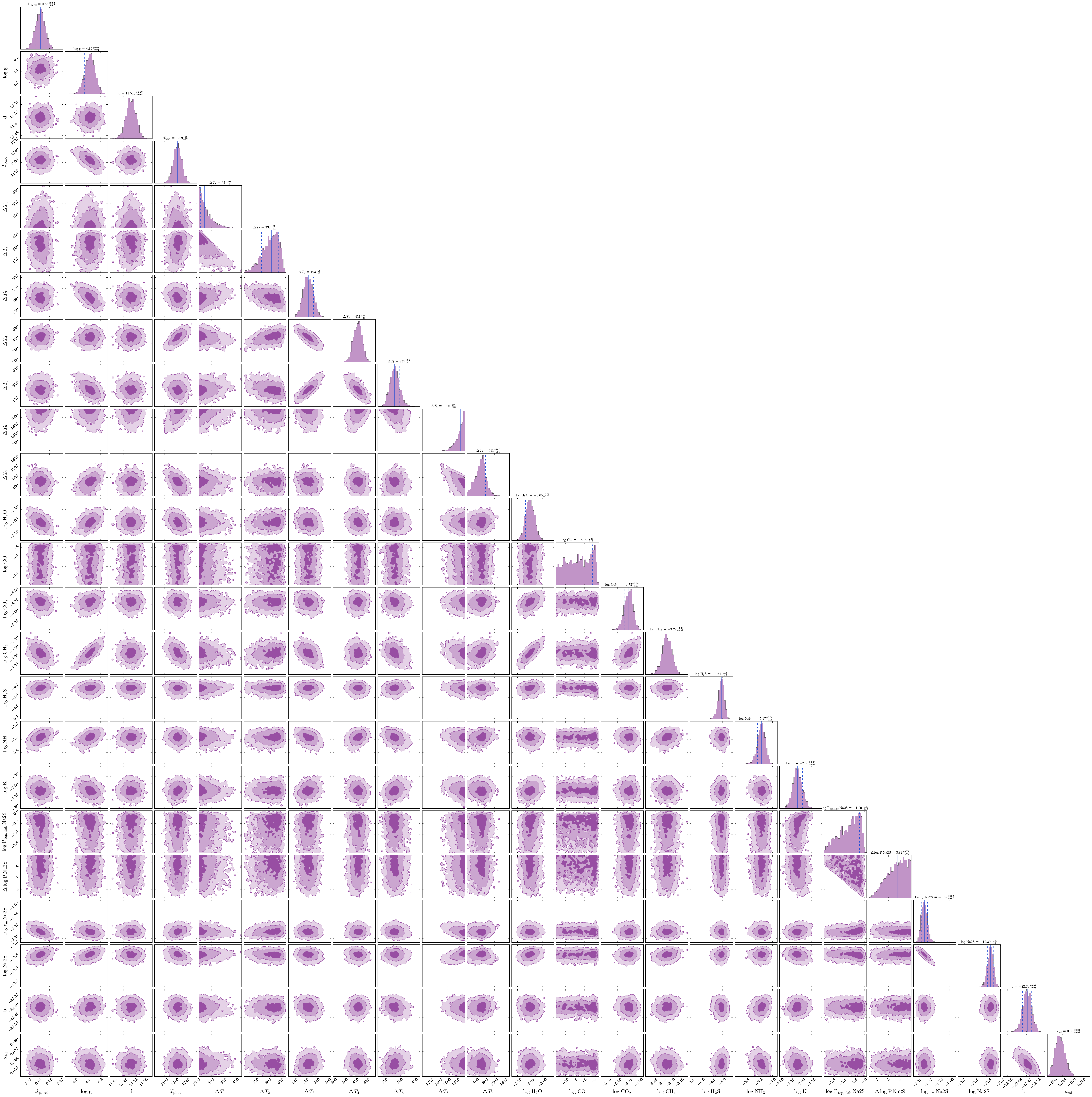}
    \caption{\change{Posterior distribution of the maximum-likelihood retrieval model, using Na$_2$S slab clouds. A full description of the free parameters and priors can be found in Table \ref{tab:retrieval_priors}.}}
    \label{fig:na2s_corner}
\end{figure}

\change{In Section \ref{sec:retrievals}, we ran four retrievals on our JWST NIRSpec Fixed Slit spectrum Ross 458c between 0.8 and 3.1 $\upmu$m using slab cloud models of different compositions (Na$_2$S, crystalline MgSiO$_3$, KCl, and MnS). The free parameters for the retrievals and their ranges are summarized in Table \ref{tab:retrieval_priors} The maximum-likelihood model used Na$_2$S clouds, and we present the resulting posterior distribution in Figure \ref{fig:na2s_corner}.}

% \section{\Poseidon\ Retrieval Corner Plot} \label{app:corner}

% \begin{figure}
%     \centering
%     \includegraphics[width=1\linewidth]{POSEIDON_FINAL_Na2S_corner.pdf}
%     \caption{Corner plot of the best-fitting \Poseidon\ retrieval, using Na$_2$S clouds.}
%     \label{fig:na2s_corner}
% \end{figure}


\bibliography{bibliography}{}
\bibliographystyle{aasjournal}



\end{document}